\documentclass[twocolumn]{aastex62}
  
\usepackage{lsstdesc_macros}
\usepackage{hyperref}
\usepackage{graphicx}
\usepackage{ulem}
\usepackage{tabularx}
\usepackage{makecell}
 
\usepackage{lineno}

\received{ABC XX, 2019}
\revised{ABC XX, 2019}
\accepted{ABC XX, 2019}

\submitjournal{The Astrophysical Journal}

\usepackage[english]{babel}
\addto\extrasenglish{%
}

\graphicspath{{./}{./figures/}}
\usepackage{rotating}
\usepackage{tocbasic}
\DeclareTOCStyleEntry[
  beforeskip=.1em plus 0.1pt,
  pagenumberformat=\textbf
]{tocline}{section}

\newcolumntype{b}{X}
\newcolumntype{s}{>{\hsize=.25\hsize}X}

\begin{document}
\title{Shear measurement bias due to spatially varying spectral energy distributions in galaxies}

\correspondingauthor{Sowmya Kamath}
\email{sowmyakth19@gmail.com}

\author[0000-0003-0443-8221]{Sowmya Kamath}
\affiliation{Kavli Institute for Particle Astrophysics and Cosmology (KIPAC)\\
2575 Sand Hill Road,\\
Menlo Park, CA 94025, USA}
\affiliation{Department of Physics, Stanford University\\
382 Via Pueblo Mall,\\
Stanford, CA 94305-4013, USA}

\author{Joshua E.\,Meyers}
\affiliation{Lawrence Livermore National Laboratory\\
7000 East Ave,\\
Livermore, CA 94550, USA}

\author{Patricia R.\,Burchat}
\affiliation{Kavli Institute for Particle Astrophysics and Cosmology (KIPAC)\\
2575 Sand Hill Road,\\
Menlo Park, CA 94025, USA}
\affiliation{Department of Physics, Stanford University\\
382 Via Pueblo Mall,\\
Stanford, CA 94305-4013, USA}

\collaboration{(LSST Dark Energy Science Collaboration)}

\begin{abstract}
Galaxy color gradients -- i.e., spectral energy distributions that vary across the galaxy profile -- will impact galaxy shape measurements when the modeled point spread function (PSF) corresponds to that for a galaxy with spatially uniform color.
This paper describes the techniques and results of a study of the expected impact of galaxy color gradients on weak lensing measurements with the Large Synoptic Survey Telescope (LSST) when the PSF size depends on wavelength.
The bias on cosmic shear measurements from color gradients is computed both for parametric bulge+disk galaxy simulations and for more realistic chromatic galaxy surface brightness profiles based on HST V- and I-band images in the AEGIS survey. For the parametric galaxies, and for the more realistic galaxies derived from AEGIS galaxies with sufficient SNR that color gradient bias can be isolated, the predicted multiplicative shear biases due to color gradients are found to be at least a factor of 2 below the LSST full-depth requirement on the {\em total} systematic uncertainty in the redshift-dependent shear calibration.
The analysis code and data products are publicly available.\footnote{https://github.com/sowmyakth/measure\_cg\_bias}
\end{abstract}


\tableofcontents
\section{Introduction}
\label{sec:intro} 

Weak gravitational lensing produces coherent alignment of images of distant galaxies due to the intervening tidal gravitational field \citep{2001PhR...340..291B}. Correlations in these distortions can be used to determine the statistical properties of the intervening mass distribution as a function of redshift. These perturbations are small when compared to the initially random orientations of galaxies, requiring a large number of galaxies for meaningful cosmological measurements. While the Large Synoptic Survey Telescope (LSST) survey, encompassing billions of galaxies, will dramatically improve the statistical power of weak lensing observations, the systematic errors must be carefully examined and controlled \citep{ScienceBook, v1DESC-SRD}.

In addition to cosmic shear, the observed shapes of galaxies are distorted by the atmosphere, the telescope optics, and sensor effects -- all of which contribute to the point spread function (PSF).
The observed images of the galaxies are thus a convolution of the true image and the observing PSF.
This makes it vital for weak lensing measurements to model the convolving PSF and correct for it using stars. 
Additional complexity arises when the distortions caused by the PSF depend on the observing wavelength \citep{2010MNRAS.405..494C}. The effects of chromatic atmospheric PSFs on weak lensing measurements have been studied by \cite{2015ApJ...807..182M} who conclude that while the predicted shear bias is significant compared to the LSST requirements, the bias can be reduced sufficiently to meet the LSST requirements if the PSF is corrected using multi-band photometry and machine learning techniques.
However, these corrections are based on the assumption that the spatial and wavelength dependencies of the galaxy surface brightness profile are separable. If instead the galaxy has a spectral energy distribution (SED) that varies across its profile -- a ``color gradient'' -- then the distortions due to the chromatic PSF will be different for different points on the galaxy. PSF corrections that are applied assuming a position-\textit{independent} SED do not correct for potential color gradient effects. Thus, it is necessary to quantify the size of this bias compared to the LSST requirements on bias for shear estimators.

Previous studies of shear bias due to color gradients have focused on the space-based Euclid survey \citep{2012SPIE.8442E..0ZA}, where the wavelength dependence of the PSF is dominated by diffraction, and the wavelength range for the (single) optical filter is $\approx$ 550\,nm to 900\,nm.
These studies have concluded that shear bias due to color gradients in surveys like Euclid can be significant compared to the requirements on systematic errors \citep{2012MNRAS.421.1385V, 2013MNRAS.432.2385S, 2018MNRAS.476.5645E}. 
We use and extend a method similar to \cite{2013MNRAS.432.2385S} (hereafter S13) to estimate the impact of color gradient (CG) bias on weak lensing measurements with the LSST, where the wavelength dependence of the PSF is dominated by atmospheric effects, and the wavelength range for each filter is $\approx 150$\,nm.

We first use a reference galaxy with extreme color gradients to demonstrate the presence of CG bias as well as its dependence on atmospheric seeing and shape measurement algorithms. We then measure the expected size of this bias on LSST weak lensing measurements by extending the study to a more representative sampling of color gradients. This is done with two different approaches, each with advantages and disadvantages:
\begin{enumerate}[label=(\alph*)]
\item We assemble a parametric catalog of galaxies with a range of color gradients that could be observed with LSST.
This method allows us to simulate galaxy images with infinite signal-to-noise ratio and measure bias from CG only. However, the accuracy of the prediction for real surveys depends on the extent to which the simulation represents color gradients in real galaxies. 
\item We use observed high-resolution galaxy images to estimate color gradients. This method allows us to leverage information from real galaxies, potentially leading to more realistic bias estimates. 
However, when the input galaxy images are noisy (as they are for real galaxies), the effect of noise cannot be removed completely and the estimated bias is a combination of CG bias and residual noise bias (see \cite{2012MNRAS.425.1951R}, for example, for details on impact of noise bias in shear measurements). 
\end{enumerate}

The parametric galaxy simulations for method (a) are generated from the LSST Catalog Simulator, \textsc{CatSim} \citep{2014SPIE.9150E..14C}, which contains astrophysical sources with properties that are representative of what the LSST will observe at its ten-year coadded depth\footnote{https://www.lsst.org/scientists/simulations}. Real galaxy images for method (b) were obtained from HST V- and I-band observations in the All-wavelength Extended Groth strip International Survey (AEGIS)\footnote{http://aegis.ucolick.org}. 

The work is presented as follows.
In \secref{explain_bias}, we describe the sources of CG bias and discuss how the galaxy's intrinsic properties, atmospheric seeing, and shape measurement algorithms affect the observed bias. 
In \secref{quant_bias}, we describe a method to isolate CG bias for a galaxy with color gradient by comparing its shear response to that of an ``equivalent" galaxy without color gradients.
In \secref{cg_para}, we apply this method to the reference galaxy with extreme color gradients and to parametric simulations from \textsc{CatSim} (approach (a)) for a more realistic estimate.
In \secref{cg_real}, we apply the method to the real HST galaxy images as seen by LSST (approach (b)).
We then summarize the results and their significance. 

To help the reader follow the description of the analysis, we compile in Tables~\ref{tab:acronyms} and \ref{tab:symbols} the acronyms and symbols used in this paper, along with the section in which they are defined in the context of this study. 

\begin{table*}[t]
\caption{Acronyms frequently used in this paper.}
\label{tab:acronyms}
\centering
\begin{tabularx}{\textwidth} {lll}
\hline
\hline
Acronym   & Description  & More details  \\
\hline
PSF    &  Point spread function: Response of imaging system to a point source.                         & \secref{psf}  \\
SED    &  Spectral energy distribution: Energy emitted as a function of wavelength.  & \secref{eff_psf} \\
CG     &  Color gradient: Varying SED across spatial profile. 							  & \secref{explain_bias}  \\
SBP    &  Surface brightness profile: Luminosity of a galaxy as a function of position and wavelength.  & \secref{galcg}  \\
HLR    & Half-light radius: Radius within which half the galaxy flux is contained.                   &  \secref{ref_param}\\
FWHM    & Full width at half maximum: For circularly symmetric SBP, diameter of circle & \\
        & at which the surface brightness is half the peak surface brightness.  & \secref{obs_cond}\\
\textsc{CatSim} &  Catalog Simulator: Catalog managed by the LSST Systems Engineering group.                               & \secref{catsim}\\
CRG    &  Galaxies simulated with the {\tt ChromaticRealGalaxy} module in {\tt GalSim}.                        & \secref{crg}  \\
SNR    & Signal-to-noise ratio: Ratio of measured flux to uncertainty on the flux. & \secref{limitations}\\
\hline
\end{tabularx}
\end{table*}

\begin{table*}[t]
\caption{Definitions of symbols used in this paper.}
\label{tab:symbols}
\begin{center}
\begin{tabularx}{\textwidth}{lll}
\hline
\hline
Symbol    &  Description  & More details  \\
\hline
$f(\vec{x},\lambda)$    &  Chromatic surface brightness profile (SBP)  & \secref{eff_psf}  \\
$\Pi(\vec{x},\lambda)$    &  Chromatic PSF  & \secref{psf}  \\
$T(\lambda)$    & Band transmission function  & \eqnref{iobs}  \\
$I^{\rm obs}(\vec{x})$    & Observed galaxy image  & \eqnref{iobs}  \\
$I^o(\vec{x})$    & PSF-corrected galaxy image  & \eqnref{io_true}  \\
$S_{\rm eff}(\lambda)$    &  Effective SED (spatial integral $f(\vec{x},\lambda)$, as a function of wavelength)   & \eqnref{S_eff} \\
$\Pi_{\rm eff}(\vec{x})$    &  Effective PSF (achromatic)  & \eqnref{psf_eff} \\
$\sigma_{\rm PSF}(\lambda)$    &  Chromatic PSF size: Radius containing 68\% of the flux for a circular Gaussian SBP & \eqnref{psf_sig} \\
$\lambda^o$    &  Reference wavelength at which chromatic PSF size is known  & \eqnref{psf_sig} \\
$\sigma^o$    &  PSF size at reference wavelength  & \eqnref{psf_sig} \\
$\alpha$    &  Scaling exponent governing PSF chromaticity & \eqnref{psf_sig} \\
$e = e_1 + {\rm i} e_2$    &  Complex ellipticity spinor of sheared galaxy shape     & \secref{weight} \\
$S(\lambda)$    &  Spectral energy distribution (SED)  & \eqnref{gal_nocg}  \\
$a(\vec{x})$    &  Spatial component of separable galaxy SBP & \eqnref{sbp}  \\
$f_{\rm CG}(\vec{x},\lambda)$    &  SBP of galaxy with color gradient  & {Equations~\ref{eqn:sbp}~and~\ref{eqn:crg_sbp}} \\
$f_{\rm no\,CG}(\vec{x})$    &  SBP of equivalent galaxy with no color gradient  & \secref{eff_psf}  \\
$e^{\rm intr}$    &  Intrinsic shape of galaxy before shear    & \secref{shape}\\
$g = g_1 + {\rm i} g_2$    &  Complex reduced shear    & \secref{shape}\\
$\hat{g}_{\rm CG}$    &  Shear estimated from galaxy with CG    & \secref{iso}\\
$\hat{g}_{\rm no\,CG}$    &  Shear estimated from equivalent galaxy with no CG    & \secref{iso}\\
$m_{\rm CG}$, $c_{\rm CG}$    &  Multiplicative and additive bias in shear estimation from CG    & \secref{iso}\\
$m_{o}$, $c_{o}$    &  Multiplicative and additive bias in shear estimation from all sources except CG    & \secref{iso}\\
$a_b(\vec{x})$, $a_d(\vec{x})$    &  Spatial components of simulated SBP of galaxy bulge and disk   & \secref{iso}\\
\hline
\end{tabularx}
\end{center}
\end{table*}

\section{Origin of color gradient (CG) bias}
\label{sec:explain_bias}
Shear measurements require accurate estimates of galaxy shapes corrected for PSF distortions.
Galaxy color gradients can produce errors in shape measurements if the following conditions hold: 
\begin{enumerate}
\item The PSF is wavelength dependent.
\item The PSF correction is performed assuming that the galaxy has a uniform SED across its profile.
\item A spatial weight function is used in the estimation of shapes.
\end{enumerate}

We explore in detail below how the three conditions produce CG bias. In this study, we restrict ourselves to analyzing the effects of CG bias for moment-based shape measurement algorithms that use weight functions matched to galaxy profiles. 
However, color gradients have also been shown to lead to shape measurement errors when ``fitting methods" are used, as described in \cite{2012MNRAS.421.1385V} where the profiles themselves weight the different regions of the image. Both \cite{2012MNRAS.421.1385V}, using fitting methods, and \cite{2013MNRAS.432.2385S}, using moment-based shape estimation, showed that the CG bias could be substantial in the Euclid survey -- exceeding nominal requirements for the multiplicative bias in the shear.

\subsection{Chromatic PSF}
\label{sec:psf}
For this analysis we approximate a chromatic PSF, $\Pi(\vec{x},\lambda)$, by a Gaussian profile with a wavelength-dependent size:
\begin{equation}
\label{eqn:psf_sig}
\sigma_{\rm PSF}(\lambda)=\sigma^o\bigg(\frac{\lambda}{\lambda^o}\bigg)^\alpha.
\end{equation}
The scaling exponent $\alpha$ is determined by the origin of the dominant source of the chromatic PSF. In the case of space-based telescopes (\eg, HST, Euclid, WFIRST), the PSF chromaticity arises primarily from diffraction due to the finite aperture: $\alpha \approx +1.0$ \citep{2010MNRAS.405..494C}. For ground-based telescopes like the LSST, the chromaticity is primarily determined by the Kolmogorov turbulence in the atmosphere: $\alpha \approx -0.2$ \citep{1966JOSA...56.1372F}. The PSF size at a reference wavelength $\lambda^o$ is defined as $\sigma_o$.

\subsection{Effective PSF for correction}
\label{sec:eff_psf}
The observed image of the galaxy, $I_{\rm CG}^{\rm obs}(\vec{x})$, can be described as the convolution (denoted by $*$) of the galaxy profile $f_{\rm CG}(\vec{x}, \lambda)$ with the chromatic PSF $\Pi(\vec{x}, \lambda)$, weighted by the band transmission function $T(\lambda)$ (which includes contributions from the atmosphere, optics, filter, and sensor) and integrated over wavelength:
\begin{equation}
\label{eqn:iobs}
I_{\rm CG}^{\rm obs}(\vec{x}) = \int f_{\rm CG}(\vec{x}, \lambda) * \Pi(\vec{x}, \lambda) T(\lambda) \, d\lambda.
\end{equation}
Since the PSF is chromatic, its effect on the observed galaxy shape will depend on the observing wavelengths. Similarly, any PSF correction to be applied to a galaxy shape will require knowledge of the galaxy SED over the observing band. If color gradients are ignored then the PSF correction will assume an ``effective'' SED that is uniform across the galaxy profile in place of the true SED of the galaxy that is position dependent. The effective SED $S_{\rm eff}(\lambda)$ is defined as the spatially integrated flux of the galaxy as a function of wavelength: 
\begin{equation}
S_{\rm eff}(\lambda) = \int f_{\rm CG}(\vec{x}, \lambda) \, d\vec{x}.
\label{eqn:S_eff}
\end{equation}
The PSF for correction will also be an ``effective" PSF $\Pi_{\rm eff}(\vec{x})$ corresponding to the image of a point source with SED $S_{\rm eff}(\lambda)$: 
\begin{equation}
\Pi_{\rm eff}(\vec{x}) = \int S_{\rm eff}(\lambda)\Pi(\vec{x}, \lambda)T(\lambda) \, d\lambda,
\label{eqn:psf_eff}
\end{equation}
where we have implicitly evaluated the convolution with a delta-function point source.
\cite{2013MNRAS.432.2385S} pointed out that color gradients do not produce a bias if the integrals are allowed to extend to infinity in each direction, which is impossible in practice. All realistic methods use some kind of spatial weighting (even if only to impose zero weight beyond the region of the ``postage stamp"), which means that different regions of the galaxy with different colors and PSFs are weighted differently, introducing a potential bias.

\subsection{Weight function in shape measurements}
\label{sec:weight}
The galaxy shapes can be expressed as complex ellipticities {\bf $e$} computed using quadrapole moments \citep{1997A&A...318..687S}:
\begin{equation}
\label{eqn:e}
e_1+{\rm i}e_2=\frac{Q_{11}-Q_{22}+2{\rm i}Q_{12}}{Q_{11}+Q_{22}+2(Q_{11}Q_{22}-Q_{12}^2)^{1/2}}\,,
\end{equation}
where $Q_{\rm ij}$ are the second moments of the galaxy image. 
 
When computing second moments, pixel values at larger distances from the centroid have more impact on the measurement than pixel values close to the centroid. 
In real galaxy shape measurements, the moments are computed from noisy images, where the noise will dominate the pixel values \citep{2011MNRAS.412.1552M}. 
Thus in order to prevent noise divergence in the calculation of second moments, it is common practice to employ weight functions of finite width to limit the integration. A weight function whose centroid, size, and ellipticity are matched to the source galaxy image optimizes the significance of the measurement. A Gaussian weight is generally preferred due to its rapid convergence to zero at large radii and the absence of singularities, as well as general mathematical convenience \citep{2003MNRAS.343..459H}.
 
By its very nature, the weight function gives more significance to the central region of the galaxy. 
For a galaxy with a spatially independent SED,
this does not produce an error as long as one correctly accounts for the weight function in the shape estimate. However, if there exists a spatial dependence of the galaxy SED, regions with different colors are weighted differently, potentially resulting in a bias in shape estimation that depends on the size and profile of the weight function and the color gradient.

\subsection{Illustration of color gradient bias for a simple bulge + disk galaxy}
\label{sec:cartoon_example}
\begin{figure*}[!ht]
\centering\includegraphics[width=0.7\linewidth]{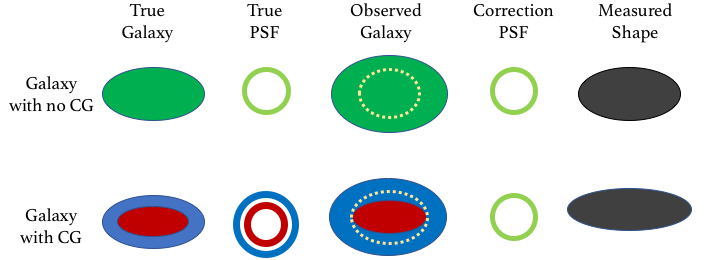}
\caption{Schematic depicting error in the measured shape of a galaxy with color gradients caused by wavelength-dependent PSF combined with a weight function for computing moments.}
\label{fig:ref_gal_show_cg}
\end{figure*}

We illustrate in \figref{ref_gal_show_cg} the impact of the aforementioned three conditions by comparing the measured shape of a galaxy with and without color gradients.
The top row depicts this measurement for a galaxy with no CG when observed with a chromatic PSF and with moments computed using a weight function.
The true galaxy (green ellipse) is convolved by the PSF (green circle) and integrated over the observing bandpass to produce the observed galaxy image. The radius of the PSF circle is proportional to the PSF size. As a result of convolution with the PSF, the observed galaxy image appears rounder and larger than the true galaxy. To obtain the correct measured shape the effect of the PSF must be measured. In the absence of CG, the effective SED is the same as the uniform galaxy SED and the effective PSF correctly encapsulates the PSF distortions. The dotted yellow circle denotes the weight function for computing moments, which are corrected using the effective PSF in order to retrieve the correct measured shape (grey ellipse).

The bottom row illustrates the same process but for a galaxy with color gradients. For simplicity we demonstrate this for a galaxy with radially dependent color gradient, where the galaxy has a compact central bulge (red ellipse) and an extended disk (blue ellipse). Both the bulge and the disk have the same centroid and shape, but have different size and color with the bulge being smaller and redder than the disk. Since the size of the PSF is color dependent, the distortion caused by the PSF on the two components is different. We assume here that the PSF is smaller for longer wavelengths (as expected for the LSST). Therefore the bulge PSF (red circle) is smaller than the PSF acting on the disk (blue circle). The observed image will be the sum of the convolution of the two components with their respective PSFs. Since the disk PSF is larger, it will have a stronger ``rounding" effect on the disk component in comparison to the smaller bulge PSF, making the disk rounder than the bulge in the observed image. The effective SED is an average of the disk and bulge SED, resulting in the effective PSF having a size in between the bulge and disk PSFs. 

As before, the weight function is the dotted yellow circle. Since the weight function effectively gives more importance to the center, the bulge will be weighted more than the disk in the shape measurement. The correction PSF (effective PSF) is larger than the actual PSF that acted on the bulge. The PSF correction step will thus overestimate the ellipticity of the galaxy.

The extent of this error in the measured shape depends on the intrinsic color gradient of the galaxy. Color gradient depends on galaxy morphology, which is correlated with redshift \citep{2010AJ....140.1528L, 2016MNRAS.460.3458K}. Therefore, color gradients can lead to redshift-dependent shape measurement biases.

\section{Isolating and quantifying CG bias}
\label{sec:quant_bias}
Shape measurement algorithms can lead to a range of different shape measurement errors that depend on noise, galaxy properties, etc.; see \cite{2018MNRAS.481.3170M}. 
To isolate color gradient bias, we use a technique described in S13: compare the shear measured from a galaxy with CG to that of an ``equivalent" galaxy with no CG. 

\begin{figure*}[ht!]
\centering\includegraphics[width=.9\linewidth]{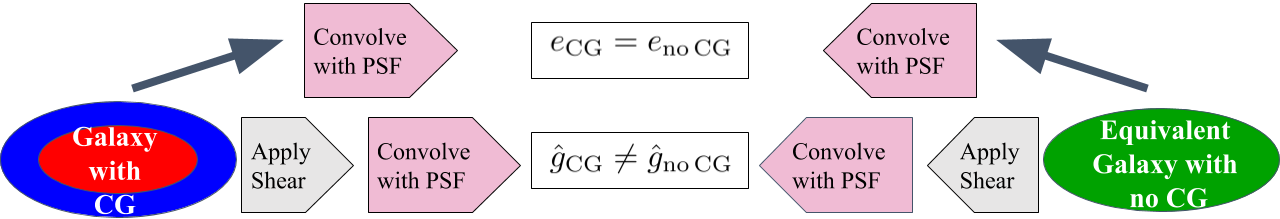}
\caption{Illustration of technique for isolating shear measurement bias due to only color gradients (CG). 
Top row: PSF-convolved ``equivalent" galaxy with no CG (right) has the same shape as PSF-convolved galaxy with CG (left). 
Bottom row: In the presence of an applied shear, the measured shapes are no longer equal, isolating CG shear bias.}
\label{fig:cg_eq_gal_flow}
\end{figure*}

The technique is illustrated in \figref{cg_eq_gal_flow}. We create a pair of galaxies -- one galaxy with a spatially dependent SED (on the left), and an equivalent galaxy with uniform SED (on the right). The SED and surface brightness profile (SBP) of the equivalent galaxy is chosen so that the two galaxies appear identical when convolved with the same chromatic PSF and observed through the same filter (top row). This leads to all non-CG biases being the same for the two galaxies.
As illustrated in the bottom row in the figure, if we apply the same shear to each galaxy and
convolve it with the same chromatic PSF, the measured shear is no longer the same. We quantify the CG shear bias as this difference between the measured shear estimators (ellipticities).
More specifically, in the absence of shear although the observed image of the galaxy with CG and the equivalent galaxy are the same, the SBP for the two galaxies are not identical. Therefore, the response to the same shear $g$ will be different for the two galaxies, causing the PSF-convolved images and the shear measured from the two observed images to be different. The difference in the two shear values ($\hat{g}_{\rm CG}$ and $\hat{g}_{\rm no\ CG}$) is an estimate of the CG bias.

\subsection{Isolating shear bias due to CG}
\label{sec:iso}
When a small shear $g$ is applied to a galaxy with no CG, the measured shear, $\hat{g}_{\rm no\,CG}$, can be approximated as
\begin{equation}
\hat{g}_{\rm no\,CG} = (1+m_o)g + c_o,
\end{equation}
where $m_o$ and $c_o$ are the multiplicative and additive bias terms.
We assume that when the same small shear $g$ is applied to a galaxy with CG, the bias on the measured shear, $\hat{g}_{\rm CG}$, due to color gradients can be encapsulated by two new terms $m_{\rm CG}$ and $c_{\rm CG}$:

\begin{equation}
\label{eqn:it_all_comes_to_this}
\hat{g}_{\rm CG}=(1+m_o+m_{\rm CG})g + (c_o + c_{\rm CG}).
\end{equation}

If the galaxy with no CG is the equivalent galaxy described earlier then taking the difference of the two measured shears will remove all other systematic bias contributions isolating contributions from color gradients only:
\begin{equation}
\label{eqn:cg}
\Delta g=\hat{g}_{\rm CG} - \hat{g}_{\rm no\,CG} = m_{\rm CG}g + c_{\rm CG}.
\end{equation}

\subsection{LSST requirements on shear calibration}
\label{sec:LSST_requirements}

As documented in secs.\,5.2, D2.1 and D2.3 in version 1 of the 
LSST Dark Energy Science Collaboration (DESC) Science Requirements Document (SRD) \citep{v1DESC-SRD}, the requirement on the total systematic uncertainty in the redshift-dependent shear calibration is that it not exceed 0.003. Combined with other requirements in the LSST DESC SRD, LSST will then be able to achieve its design constraints on dark energy \citep{v1DESC-SRD, 2006astro.ph..9591A}.

The SRD analysis uses five photometric redshift bins, each with a width of $\Delta z=0.2$, in the redshift range $0.2 \leq z \leq 1.2$.
A linear parameterization is suggested for the redshift dependence of the multiplicative shear bias, $m(z)$: 
\begin{equation}
    m\left(z\right) = m_{z} \left(\frac{2z - z_{\max}}{z_{\max}} \right) + m_{\rm avg}\,,
\label{eqn:tomo-z-equation}
\end{equation}
where $z_{\max}$ is the redshift value at the center of the highest redshift bin, 
$m_{\rm avg} = m(z_{\max}/2)$ is the average value of $m$ in the range $z \in \left[0,z_{\max}\right]$, and 
$2m_{z} = m(z_{\max}) - m(0)$ is the total variation in $m$ in the range $z \in \left[0,z_{\max}\right]$.
The 0.003 requirement is then on the total systematic uncertainty in the redshift-dependent shear calibration $m_z$ due to all contributions to multiplicative shear bias.
In principle, there is some higher-order dependence on the redshift-dependent function adopted for $m(z)$. 
In the absence of an explicit requirement for the case of {\em non-linear} dependence of $m(z)$ on $z$, we map the requirement of $2(0.003) = 0.006$ on the uncertainty on $2m_{z} = m(z_{\max}) - m(0)$ onto a requirement on the uncertainty on the maximum span of $m(z)$ over the redshift range [0, 1.2].

\subsection{Creating equivalent galaxy with no color gradient}
\label{sec:make_eq}
The method described in \secref{iso} for isolating CG bias assumes we can create an equivalent galaxy with no CG with the properties shown in \figref{cg_eq_gal_flow}. We describe below a method adapted from Section 2 of S13 to create an equivalent galaxy with no CG.

In the absence of color gradients, the galaxy SBP can be factored into a product of spatial and spectral components $a(\vec{x})$ and $S(\lambda)$:
\begin{equation}
\label{eqn:f_nocg}
f(\vec{x}, \lambda) = a(\vec{x})\, S(\lambda).
\end{equation}
The true image of the galaxy, $I^{o}(\vec{x})$, through the transmission function $T(\lambda)$ can be written as
\begin{equation}
\label{eqn:io_true}
I^{o}(\vec{x}) = \int f(\vec{x},\lambda) \, T(\lambda) \, d\lambda.
\end{equation}
However, as shown in \eqnref{iobs}, the observed galaxy is a convolution of the true galaxy and the PSF, in the observed band:
\begin{equation}
I^{\rm obs}(\vec{x}) = \int f(\vec{x},\lambda) * \Pi(\vec{x},\lambda) \, T(\lambda) \, d\lambda.
\end{equation}
In Fourier space, convolution becomes multiplication and the expression for the observed galaxy image becomes
\begin{eqnarray}
\label{eqn:iob_f}
  \tilde{I}^{\rm obs}(\vec{k}) &= &\int \tilde{a}(\vec{k}) \, S(\lambda) \, \tilde{\Pi}(\vec{k},\lambda) \, T(\lambda) \, d\lambda \nonumber \\
  &=& \tilde{a}(\vec{k}) \int S(\lambda) \, \tilde{\Pi}(\vec{k},\lambda) \,  T(\lambda) \, d\lambda,
\end{eqnarray}
while the unconvolved image in Fourier space is
\begin{eqnarray}
\label{eqn:io_f}
    \tilde{I}^{\rm o}(\vec{k}) &= & \int  \tilde{f}(\vec{k}, \lambda) \,  T(\lambda) \, d\lambda \nonumber \\
    &=& \int\tilde{a}(\vec{k}) \, S(\lambda) \,  T(\lambda) \, d\lambda.
\end{eqnarray}
Solving \eqnref{iob_f} for $\tilde{a}(\vec{k})$ and substituting in \eqnref{io_f} we get
\begin{eqnarray}
\tilde{I}^o(\vec{k}) &= & \int  \frac{\tilde{I}^{\rm obs}(\vec{k})}{\int S(\lambda) \,  \tilde{\Pi}(\vec{k},\lambda) \, T(\lambda) d\lambda}S(\lambda) \, T(\lambda) \, d\lambda. 
\end{eqnarray}
The denominator is the effective wavelength-independent PSF, $\Pi^{\rm eff}(\vec{x})$, computed in \eqnref{psf_eff}, where the effective SED is identical to the uniform galaxy SED. Thus, the SBP of the galaxy at a particular wavelength $\lambda_{\rm ref}$ can be written as
\begin{equation}
\label{eqn:sbp_iobs}
\tilde{f}^o(\vec{k},\lambda_{\rm ref})= \frac{S(\lambda_{\rm ref}) \, \tilde{I}^{\rm obs}(\vec{k})}{\tilde{\Pi}^{\rm eff}(\vec{k})}.
\end{equation}
In other words, in the absence of color gradients the galaxy SBP at a given wavelength is the PSF convolved galaxy image, deconvolved by the effective PSF.
 
\begin{figure*}[!htpb]
    \includegraphics[width=1\linewidth]{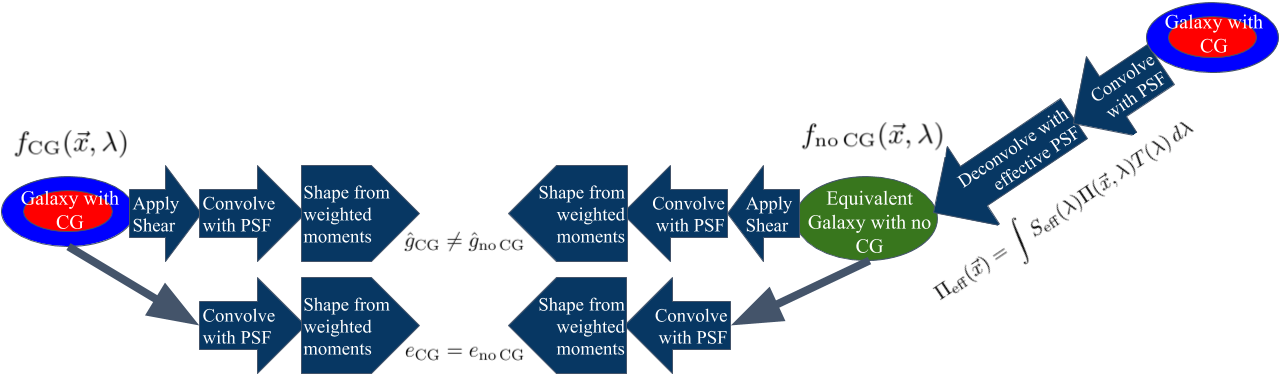}
 \caption{Schematic showing the complete process of creating an equivalent galaxy with no CG (green ellipse) from a galaxy with CG (multi-colored ellipse). See \secref{make_eq} in text for details.}
 \label{fig:cartoon_whole}
\end{figure*}

We now extend the above analysis to a galaxy with CG. We approximate the SBP of the equivalent galaxy as \eqnref{sbp_iobs} applied to the observed image of a galaxy with CG with the same effective SED but no color gradient. Thus the SBP of the effective galaxy with no CG is
\begin{eqnarray}
\label{eqn:gal_nocg}
\tilde{f}_{\rm no \,CG}(\vec{k},\lambda)&=&\frac{S_{\rm eff}(\lambda)\tilde{I}_{\rm CG}^{\rm obs}(\vec{k})}{\tilde{\Pi}_{\rm eff}(\vec{k})},
\end{eqnarray}
where $S_{\rm eff}(\lambda)$ is the effective SED computed in \eqnref{S_eff}. In the absence of an applied shear, the observed image corresponding to a galaxy with SBP of \eqnref{gal_nocg} is identical to $\tilde{I}_{\rm CG}^{\rm obs}(\vec{k})$. A flowchart illustrating the entire method is also shown in \figref{cartoon_whole}.

We use the modular galaxy simulation toolkit {\tt GalSim}\footnote{https://github.com/GalSim-developers/GalSim} \citep{2015A&C....10..121R} to perform the integrations, convolutions and deconvolutions described above to simulate both the galaxy with CG and the equivalent galaxy with no CG. While the galaxy image with CG, $I_{\rm CG}^{\rm obs}(\vec{x})$ in \eqnref{iobs}, is drawn\footnote{Draw refers to the {\tt GalSim} {\tt drawImage} function, which draws an object profile onto an image.} with the LSST pixel scale, the images in Fourier space, $\tilde{I}_{\rm CG}^{\rm obs}(\vec{k})$ and $\tilde{\Pi}_{\rm eff}(\vec{k})$ in \eqnref{gal_nocg}, are drawn with four times the resolution in order to reduce errors produced by finite pixel size during deconvolution.

\subsection{Shear estimation from galaxy shapes}
\label{sec:shape}
We described how galaxy shapes are measured from galaxy images in \secref{weight}, quantifying the ellipticities from the second moments with \eqnref{e}. In this section we describe the procedure used to estimate the shear from galaxy shapes. The reduced shear $g=g_1+{\rm i}g_2$ applied to an unlensed source with intrinsic shape $e^{\rm intr}$ is related to the observed ellipticity $e$ by the transformation \citep{1997A&A...318..687S}
\begin{equation}
e = \frac{e^{\rm intr} + g}{1+g^*e^{\rm intr}}, \quad {\rm if} \, |g|\leq 1.
\end{equation}

Assuming that the intrinsic orientation of the galaxies is random, the mean of their observed sheared ellipticities is the reduced shear, $\langle e\rangle = g $. Because the weak lensing shear is so small compared to the intrinsic, randomly oriented galaxy ellipticities (shape noise), averaging over very large ensembles of galaxies is necessary to achieve small statistical uncertainties \citep{2018MNRAS.481.3170M}. We employ a technique to suppress shape noise without generating a large number of galaxies by simulating six galaxies with equidistant intrinsic ellipticity on a ring around 0 -- ``ring test'' \citep{2007AJ....133.1763N} to measure shear (see, for example, section 6.1 in \cite{2015ApJ...807..182M}).
 
\subsection{Moment-based shape estimation with {\tt GalSim}}
\label{sec:moments}
{\tt GalSim} is used to perform shape estimation, in addition to the image simulation described above.
The shapes are computed with ``adaptive" moments that use an elliptical Gaussian weight function, the shape of which is matched to that of the PSF-convolved galaxy image. The measured moments of the galaxy are corrected for the PSF to produce the ellipticity of the intrinsic galaxy image \citep{2002AJ....123..583B}.

For most of this analysis, we use the {\tt GalSim} implementation of the ``re-Gaussianization" algorithm ({\tt REGAUSS}), which treats the deviations of the PSF from Gaussianity perturbatively in real space, without first measuring moments \citep{2003MNRAS.343..459H}. 

We found that, when compared to other shape estimation algorithms in {\tt GalSim}, the {\tt REGAUSS} algorithm makes more accurate shape measurements. The error in estimated shape is approximately $10^{-6}$ for a simulated Gaussian galaxy with a Gaussian PSF drawn with LSST pixel scale; the bias is larger for more complex situations.

Our method (described above) of using an equivalent galaxy without color gradients to isolate the contribution from color gradients reduces the impact of the shape estimation algorithm on the estimated CG bias. 
For the reference galaxy simulated in a redshift range of 0-1.2, we estimate the CG bias using three other adaptive moment estimation and PSF correction algorithms implemented in GalSim: {\tt KSB}\footnote{The {\tt GalSim} {\tt KSB} algorithm is a specific implementation of the KSB method \citep{1995ApJ...449..460K, 1997ApJ...475...20L}, as described in Appendix C of \cite{2003MNRAS.343..459H}.}, {\tt LINEAR} \citep{2003MNRAS.343..459H} and {\tt BJ} \citep{2002AJ....123..583B}.
The difference in CG bias estimates for {\tt REGAUSS}, {\tt KSB} and {\tt BJ} is less than $1 \times 10^{-4}$ for all redshifts. For {\tt LINEAR} and {\tt REGAUSS}, the difference is also less than $1 \times 10^{-4}$, except for $z \lesssim 0.2$, when the difference in bias is just under $4 \times 10^{-4}$.

As discussed in \secref{weight}, the weight function used in the shape measurement has a very significant effect on CG bias. 
Therefore, it is important to study the effect of the size of the weight function on the measured CG bias. The implementation of {\tt REGAUSS} in {\tt GalSim} does not have the provision to switch off or fix the size of the weight function computed with adaptive moments. 
Therefore, to study the impact of the weight function size, we use the {\tt GalSim} implementation of the {\tt KSB} shape measurement algorithm where the size of the circular weight function can be fixed by the user. 

While we found that {\tt REGAUSS} robustly yields correct results for noise-free parametric simulations and real galaxies with high signal-to-noise ratio (SNR)\footnote{We use the same definition of SNR as that used in the GREAT3 challenge \citep{2014ApJS..212....5M}, where SNR is defined as the ratio of measured flux to uncertainty on the flux within an elliptical Gaussian aperture matched to the size and shape of the PSF-convolved galaxy image.}, we found a high failure rate for low SNR galaxies. This is because the adaptive moment algorithm is not robust to noise and therefore does not always converge to a solution. This resulted in elimination of a significant fraction of small, faint galaxies from our analysis sample. Although these galaxies are not likely to be used in weak lensing analyses, we recommend that shape estimation methods that are more robust to noise be used in future CG bias studies.

\section{Estimating CG bias with parametric galaxies }
\label{sec:cg_para}
We first evaluate the impact of color gradients on shear measurements with simulated parametric galaxies. Since the galaxies can be simulated with no noise, this enables us to isolate the effect of CG bias without being contaminated by noise bias. 

\subsection{Observing conditions}
\label{sec:obs_cond}
\begin{figure}[ht]
\centering\includegraphics[width=1.\linewidth]{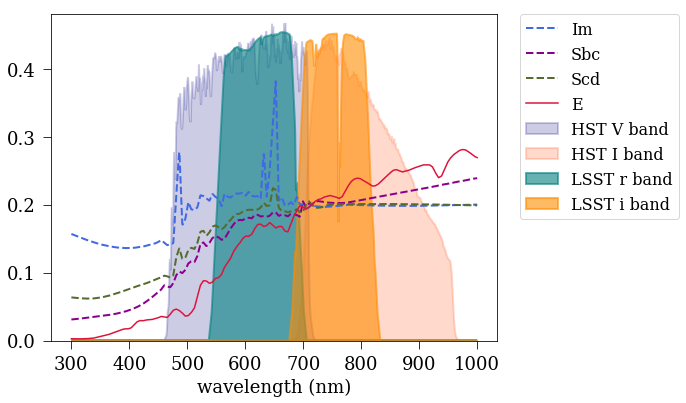}
\caption{Transmission probabilities for LSST $r$ and $i$ bands and HST V and I bands. Also shown are three types of SEDs: CWW-Im, CWW-Sbc, and CWW-Scd spectra (dashed lines), assigned as disk SEDs in the CG analysis of the reference galaxy, and CWW-E spectrum (solid line), assigned as the bulge SED. The SEDs shown here have been normalized to 0.1 at 550\,nm and redshifted to 0.3.}
\label{fig:ref_gal_filters}
\end{figure}

The galaxy images are simulated in the two main LSST lensing bands, $r$ and $i$, with the measurements performed in the $r$ band, unless otherwise noted. 
In \figref{ref_gal_filters}, we show the transmission functions for the LSST $r$ and $i$ bands\footnote{https://github.com/lsst/throughputs/tree/master/baseline}, as well as the HST V and I bands used in the analysis of AEGIS galaxies, described below.

As described in \eqnref{psf_sig}, we model the LSST PSF as a circular Gaussian profile with a wavelength-dependent size that is seeing limited with scaling exponent $\alpha=-0.2$. 
This value is valid for purely Kolmogorov turbulence -- i.e., an infinite outer scale -- in the limit of long exposure times. 
For this analysis we set the PSF size to $\sigma^o=0.297\rm \,arcsec$ (FWHM=0.7 arcsec) at $\lambda^o=550$\,nm. This value was chosen to match the expected LSST median zenith seeing in the $r$ band \citep{Overview}. 
We study the impact of the scaling exponent and PSF size on measured bias.

We limit the scope of this study to only wavelength-dependent PSF size and do not include any wavelength-dependent PSF ellipticity effects, such as atmospheric differential chromatic refraction (DCR).

\subsection{Parametric galaxy with color gradients}
\label{sec:galcg}
We model galaxies with CG as concentric bulge + disk components with different SEDs, with the following SBP:
\begin{equation}
f_{\rm CG}(\vec{x},\lambda)=a_b(\vec{x})S_b(\lambda)+a_d(\vec{x})S_d(\lambda),
\label{eqn:sbp}
\end{equation}
where $a_{b,d}(\vec{x})$ are the spatial profiles and $S_{b,d}(\lambda)$ the SEDs of the bulge and disk components denoted by the subscripts $b$ and $d$, respectively. We use this parameterization of galaxies with CG for two different samples: 1) the reference galaxy with extreme color gradients and 2) the \textsc{CatSim} catalog of galaxies.

\subsection{Reference parametric bulge + disk galaxy with extreme color gradients}
We first demonstrate the presence of CG bias for a simulated reference galaxy with extreme color gradients: a superposition of a small red bulge and an extended blue disk. The parameters of the galaxy are chosen to be identical to the ``B" type galaxy in S13, which allows us to compare results for the same galaxy model. Indeed, we find that our estimates of CG shear bias for the reference galaxy with a Euclid PSF and bandpass are in agreement with the values estimated in S13.

\subsubsection{Parameters for simulating extreme-CG galaxy}
\label{sec:ref_param}
The spatial components of the bulge and disk SBP, $a_b(\vec{x})$ and $a_d(\vec{x})$, are modelled as Sersic profiles \citep{1963BAAA....6...41S} with indices $n_s =1.5$ \footnote{We chose the bulge Sersic index to match the value in S13, rather than using the more conventional value $n_s=4$. 
We find that the predicted CG shear bias $m_{\rm CG}$ is up to 26\% smaller in magnitude when the reference galaxy bulge Sersic index is taken to be 4 rather than 1.5.} and $1.0$ for the bulge and disk, respectively. The bulge is compact with a half-light radius (HLR) of $0.17\rm \, arcsec$. 
The spatial component of the bulge accounts for 25\% of the total galaxy flux. The extended disk is much larger with HLR of $1.2 \rm \,arcsec$. Both, the bulge and disk, have the same centroid positions and ellipticities, $e^{\rm intr}=(0.3, 0.3)$.
 
The SEDs of the bulge and disk are shown in \figref{ref_gal_filters} with the redder bulge assigned a CWW-E spectrum (solid red line) and the disk a CWW-Im spectrum (dashed blue line) with stronger emission lines (see \citep{1980ApJS...43..393C}). To further investigate the impact of intrinsic galaxy color gradients, we measure the CG bias for the reference galaxy with extreme color gradients at different redshifts and for two additional disk SEDs: CWW-Sbc and CWW-Scd (dashed magenta and green lines, respectively). The SEDs in \figref{ref_gal_filters} are normalized to 0.1 at 550\,nm in the rest frame and then redshifted to 0.3. The bulge and disk parameters are summarized in \tabref{para_parameters}.
 
\begin{table*}[ht]
\caption{Parameters of simulated galaxies for bulge/disk components.}
\label{tab:para_parameters}
\begin{center}
\begin{tabular}{ccccc}
\hline
\hline
Name & Sersic index & SED & HLR (arcsec) & Flux proportion (\%)   \\
\hline
\makecell{Reference galaxy with extreme\\ color gradients} & 1.5/1 & E/(Im, Scd, Sbc) & 0.17/1.2 & 25/75\\
\textsc{CatSim} galaxies (mean) & 4/1 & Template SEDs & 0.2/0.4 & 20/80 \\
\hline
\end{tabular}
\end{center}
\textbf{Note.} Values correspond to bulge/disk components of the reference galaxy and \textsc{CatSim} galaxies. The entries for the \textsc{CatSim} galaxies are the mean bulge and disk parameters of the entire sample. The \textsc{CatSim} SEDs are drawn from an ensemble of template SEDs.
\end{table*}

\subsubsection{Verification of method to isolate CG bias}
We verify our proposed method for isolating CG bias described in \secref{quant_bias} by simulating an equivalent galaxy with no CG, for the reference galaxy, and then applying the same shear and PSF to both the original galaxy and the equivalent galaxy. 
The shears recovered with a ring test for each galaxy are compared. 
The galaxy shapes are estimated with the {\tt REGAUSS} method of {\tt GalSim}. 
Since the bias values are found to be similar for the two shear components $g_1$ and $g_2$, we report results here for only $g_1$.

\begin{figure*}[!ht]
\centering\includegraphics[width=1\linewidth]{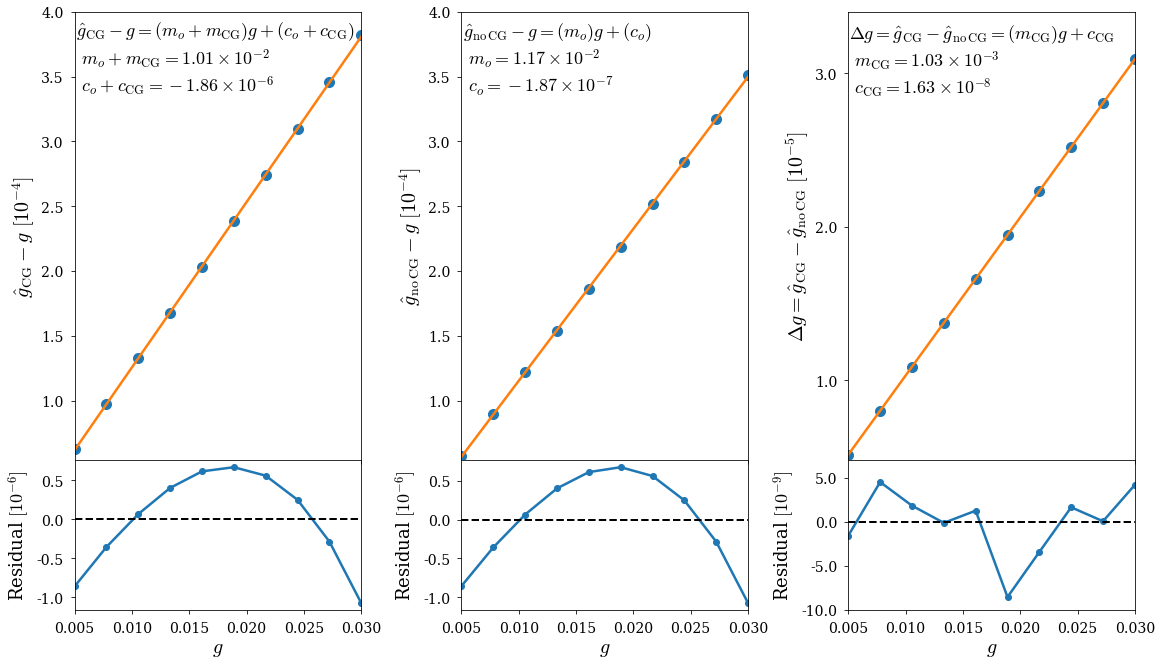}
\caption{Tests of the validity of the definition of multiplicative and additive shear biases $m_{\rm CG}$ and $c_{\rm CG}$ (\secref{iso}) for simulated LSST galaxies.
The horizontal axes correspond to applied shear $g$.
The vertical axis in the top panels corresponds to 
(left) the difference between measured and applied shear for a galaxy with CG, (center) the difference between measured and applied shear for the equivalent galaxy with no CG, and 
(right) the difference between measured shears for galaxies with and without CG. 
The blue dots correspond to these differences and the orange lines to linear fits. 
The values of the multiplicative and additive biases from the fits are shown near the top of each panel. 
In the bottom plot of each panel, we show the difference between the linear fit and the data points.}
\label{fig:lsst_check_def}
\end{figure*}

In the top left and center panels of \figref{lsst_check_def}, we plot as a function of applied shear $g$ the difference between measured shear $\hat{g}_{\rm CG}$ and applied shear for a galaxy with CG (left), and the difference between measured shear $\hat{g}_{\rm no \, CG}$ and applied shear for the equivalent galaxy with no CG (center). 
The blue dots correspond to these differences and the red line corresponds to the result of a linear fit to these values. 
The bottom plot in each panel corresponds to the difference between the linear fit and the points.
We see that for this range of applied shear ($g\leq 0.03$), the measured shear is linearly dependent on the applied shear. 
The multiplicative and additive biases, corresponding to the slope ($m$) and intercept ($c$) of the fits, are found to be $m=1.26 \times 10^{-2}$ and $c=-1.48 \times 10^{-7}$ for the original galaxy with CG, 
and $m_o=1.16 \times 10^{-2}$ and $c_o=-1.63 \times 10^{-7}$ for the equivalent galaxy with no CG.

We proposed in \secref{iso} that $m_o$ includes the shear bias from all sources except CG and that the bias from CG can be isolated as $m_{CG} = m - m_o$. 
We test this hypothesis by plotting in the right panel of \figref{lsst_check_def} the difference $\Delta g = \hat{g}_{\rm CG}- \hat{g}_{\rm no \, CG}$ as a function of applied shear. 
The small differences between the linear fit and the points, shown in the bottom right panel, validate the linear relationship assumed in \eqnref{cg}. 
The slope of the linear fit is an estimate of bias from only CG: $m_{\rm CG}=1.03 \times 10^{-3}$. The sign of $m_{\rm CG}$ is positive; \ie, the measured shear is larger than the true shear, in agreement with the example in \secref{cartoon_example}. 

We now apply this method to study the dependence of CG bias on intrinsic color gradient and observing conditions. 
Due to the small magnitudes of the additive bias in our simulations, which is expected for a circular PSF with circularly symmetric wavelength-dependence, we focus our analysis on studying the impact of color gradients on only the shear multiplicative bias.

\subsubsection{Results of CG bias measurements on extreme-CG galaxy}
To illustrate the impact of intrinsic galaxy color gradients on CG bias we estimate the CG bias for the reference galaxy with extreme color gradients. The estimated bias values here are not representative of the bias expected for shear measurements from all galaxies seen by LSST; rather, we expect these bias values to be outliers for a more representative sample.

\begin{figure*}[!htbp]
\centering\includegraphics[width=0.8\linewidth]{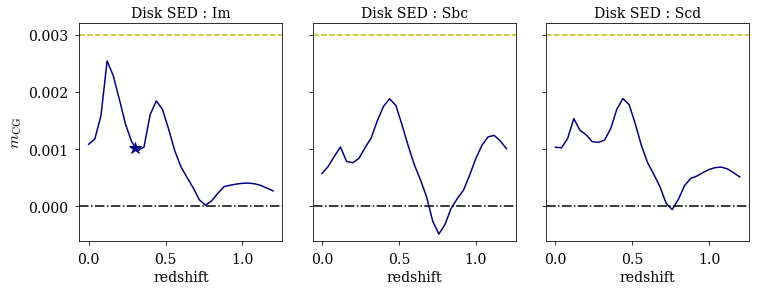}
\caption{Multiplicative shear bias due to CG, $m_{\rm CG}$, as a function of redshift for a reference galaxy with CWW-E bulge SED and three different disk SEDs: CWW-Im, CWW-Sbc, CWW-Scd (left to right). The star denotes the CG bias for the galaxy with CWW-Im disk SED at redshift of 0.3.}
\label{fig:ref_gal_diff_sed}
\end{figure*}

We plot in \figref{ref_gal_diff_sed} the estimated value of $m_{\rm CG}$ for three different disk SEDs, with the same bulge SED (E), for different redshifts. The CG biases are estimated from the shear recovered when a constant shear of $g=(0.01, 0.01)$ is applied to the galaxy.
The bias is larger for the Im disk SED, which has emission lines. 
Since CG bias arises due to the difference in bulge and disk color in the observing bandpass, galaxies with the same bulge and disk SEDs can have different CG bias at different redshifts depending on their SED profiles. 
The yellow dashed lines show the LSST DESC requirement of 0.003 on the total systematic uncertainty in the redshift-dependent shear calibration.
The estimated CG bias for all three disk SEDs and redshifts are observed to be lower than the LSST requirement, with the mean CG bias at all redshifts and disk SEDs being $m_{\rm CG} = 8.65 \times 10^{-4}$. 

In S13, the magnitude of the CG bias for the reference galaxy in the same redshift range, with Euclid's expected PSF size and wavelength dependence, was estimated to be in the range $4\times10^{-4} - 1.5\times10^{-3}$, which is not dissimilar to our predictions for LSST for this extreme case. 
The observing bandpass for Euclid covers a wider wavelength range (550nm-900nm) than those for LSST and the slope of the PSF chromaticity is steeper, leading to higher sensitivity to color gradients in Euclid.
However, the Euclid PSF is smaller (FWHM=0.15 arcsec) than the LSST PSF, offsetting the other effects.

\subsubsection{Dependence on weight function}
\begin{figure}[!ht]
\centering\includegraphics[width=.9\linewidth]{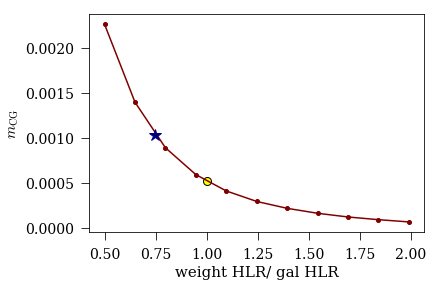}
\caption{Multiplicative shear bias from CG, $ m_{\rm CG}$, as a function of the ratio of weight function size to galaxy size (red dots) with shape measurements performed using the {\tt GalSim} implementation of KSB shape estimation algorithm. 
We show the value of the CG bias when the weight function is computed with adaptive moments (blue star) and when the weight function HLR is equal to the galaxy HLR (yellow circle).}
\label{fig:ref_gal_weight}
\end{figure}

As described in \secref{weight}, CG bias occurs in the presence of a weight function that assigns more weight to certain parts of the galaxy image. In the absence of a weight function, we expect the CG bias to be zero \citep{2013MNRAS.432.2385S}. To study the effect of our choice of weight function, we estimate CG bias while varying the size of the circular Gaussian weight function using the KSB method for galaxy shape measurement in {\tt GalSim}. The results of the analysis for the reference galaxy with Im disk SED at redshift 0.3 is shown in \figref{ref_gal_weight}. The values on the horizontal axis correspond to the ratio of the HLR of the weight function to the bulge+disk galaxy HLR (0.949 arcsec). 
As expected, the CG bias decreases with increasing size of the weight function with the bias approaching zero for large weight sizes.
The yellow circle denotes the CG bias for a Gaussian weight function with HLR equal to the size of the galaxy as used in S13. For comparison we show that the CG bias computed with REGAUSS for the reference galaxy (blue star) is larger. This is because the {\tt GalSim} REGAUSS algorithm uses adaptive moments to match an elliptical Gaussian weight to the galaxy image. Since the reference galaxy is the sum of two Sersics, the Gaussian profile matched by REGAUSS has an HLR = 0.7 arcsec, smaller than the PSF convolved galaxy HLR of 0.949 arcsec. Thus this analysis uses a smaller weight function than in S13, leading to larger biases. 
 
\subsubsection{Dependence on PSF}
\begin{figure*}[!ht]
\centering\includegraphics[width=.8\linewidth]{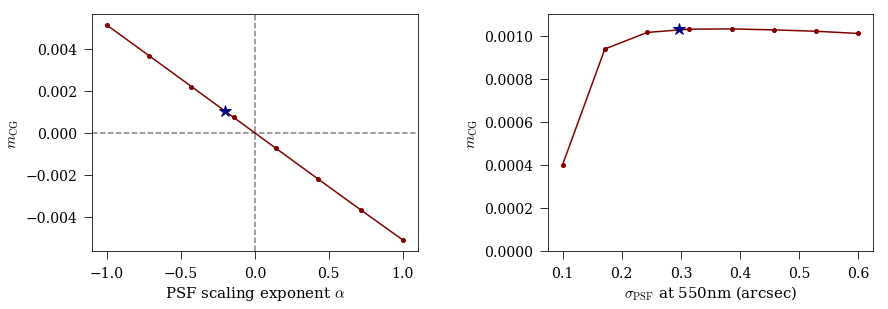}
\caption{Left: Multiplicative shear bias from CG, $m_{\rm CG}$, as a function of PSF chromatic scaling exponent $\alpha$. Right: $m_{\rm CG}$ as a function of PSF size. The stars correspond to the PSF parameters used in the reference galaxy analysis: $\alpha=0.2$ and $\sigma^o=0.297$ arcsec at $\lambda^o=550$ nm.}
\label{fig:ref_gal_psf_cg}
\end{figure*}

We illustrate the effect of the PSF chromaticity on CG bias in \figref{ref_gal_psf_cg}. The left panel shows the dependence on the PSF size scaling exponent $\alpha$. A scaling exponent of zero is an achromatic PSF and thus has no CG bias.
The sign of the exponent determines the sign of the bias, with bias being positive for a negative exponent (ground-based survey) and negative for positive exponents (space-based surveys). The CG bias is highly sensitive to the chromatic nature of the PSF, with a change of 0.1 in the value of $\alpha$ changing the CG bias by 20\% of the LSST requirement. The panel on the right in \figref{ref_gal_psf_cg} shows the dependence of CG bias on the size of the PSF. For small PSF sizes, the image distortions and PSF corrections are small, resulting in smaller CG bias. When the PSF is larger than the bulge ($\sigma \approx $ 0.15 arcsec), the CG bias comes into play and $m_{\rm CG}$ remains constant even at larger PSF sizes. The blue star is the same as in \figref{ref_gal_diff_sed} and shows the bias for the reference galaxy PSF with $\alpha=-0.2$ and $\sigma^o=0.297$\,arcsec at $\lambda^o=550\,$nm.

We also found that the predicted CG shear bias $m_{\rm CG}$ was up to 20\% smaller when the reference galaxy is convolved with a Kolmogorov PSF, rather than a Gaussian PSF, with the same full width at half-maximum value and the same wavelength-dependent scaling exponent. Since the aim of this analysis is to place conservative upper limits on CG bias estimates, we use the nominal Gaussian PSF for the rest of the analysis.

\subsection{ \textsc{CatSim} catalog of parametric galaxies}
\label{sec:catsim}

Although the CG bias for the reference galaxy computed above is small, it is not negligible in comparison to the LSST requirement for \textit{all} systematic effects. Since we also observe a large dependence of CG bias on the galaxy intrinsic properties, CG bias cannot be outright ignored. Therefore, to more accurately estimate the expected CG bias for LSST, we calculate the bias using a catalog of galaxy characteristics (redshift, size, magnitude and shape) expected in a 10-year LSST weak lensing dataset. 

We use a simulated catalog prepared by the LSST Catalog Simulator, \textsc{CatSim} \citep{2014SPIE.9150E..14C}, containing astrophysical sources with properties that are representative of what the LSST will observe to its coadded depth. The catalog covers a $4.5\times 4.5$ degree footprint on the sky with realistic galaxy morphologies, apparent colors and spatial distributions, and redshifts. The galaxy simulation is based on dark matter halos from the Millennium Simulation \citep{2005Natur.435..629S} and a semi-analytic baryon model described in \cite{2006MNRAS.366..499D}. Galaxy morphologies are modeled using two Sersic profiles, where the bulge and disk components have Sersic indices 4 and 1, respectively. 
For all sources, an SED is fit to the galaxy colors using a sophisticated fitting algorithm, independently for the bulge and disk, that includes inclination-dependent reddening.
The catalog contains galaxies with $r$-band AB magnitudes brighter than 28.

\subsubsection{Parameters of \textsc{CatSim} galaxies}
For our analysis, we select galaxies from 858,502 galaxies in the one-square-degree \textsc{CatSim} catalog.
The selected galaxies satisfy the following criteria, where the number in parentheses is the number of galaxies satisfying the cumulative criteria:
\begin{enumerate}
\item Composed of bulge and disk components -- i.e., there is a non-zero color gradient (245,359).
\item Redshift less than 1.2 (76,201).
\item $i$-band magnitude $<25.3$ (58,310). 
\item Bulge and disk semi-major axis each less than 3 arcseconds ($\approx$ 15 pixels) \footnote{We impose this maximum size to avoid {\tt GalSim} run-time errors caused by Fourier transform arrays being too large in the {\tt ChromaticRealGalaxy} modelling described later in \secref{crg}.}(57,943) .
\end{enumerate}

 The {\tt GalSim} {\tt REGAUSS} shape measurement algorithm fails for a fraction of galaxies with CG or their equivalent galaxies with no CG. 
While the CG bias estimates shown in this section are for noise-free \textsc{CatSim} simulations, we also study the impact of Poisson noise in the image on our CG bias estimates, as explained in detail in \secref{limitations} and \appref{crg_catsim}. 
 As described earlier in \secref{moments}, {\tt GalSim} shape estimation methods have a high failure rate for small noisy galaxy images. We exclude such galaxies from our final sample if the shape measurement failed for any of the six galaxy shape measurements in the ring test of the galaxy with CG or its equivalent galaxy with no CG. Our final results are based on 45,534 \textsc{CatSim} galaxies where the shape measurement was always successful.

\begin{figure*}[!ht]
\centering\includegraphics[width=.9\linewidth]{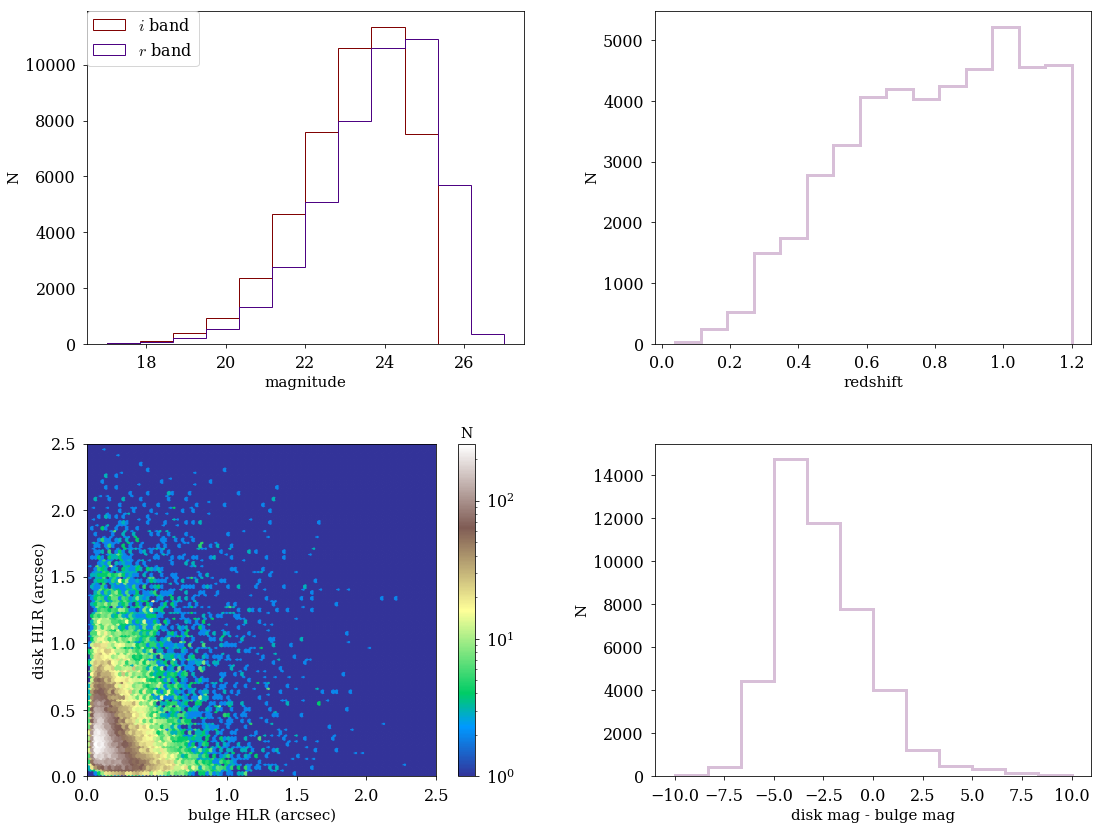}
\caption{Distributions of galaxy parameters for selected \textsc{CatSim} galaxies.
Top left: $i$- and $r$-band AB magnitudes.
Top right: Galaxy redshifts. Bottom left: Half-light radius (in arc seconds) of the bulge and disk components.
Bottom right: Difference between disk and bulge magnitudes.}
\label{fig:cat_gal}
\end{figure*}

In \figref{cat_gal} we show the distribution of the intrinsic parameters of the selected galaxies. The top-left panel shows a histogram of the galaxy AB magnitudes in the $i$ and $r$ bands. With the criteria being applied to select galaxies in the ``Gold Sample" ($i<25.3$) \citep{ScienceBook}, the sample includes galaxies with $r$-band magnitudes up to 26.5. The top-right plot shows the distribution of galaxy redshifts. The bottom-left panel shows the distribution of bulge and disk sizes, with mean HLR of 0.2 and 0.4 arcsec, respectively. A large fraction of the selected galaxies are small, with a large number of galaxies with HLR comparable to an LSST pixel (0.2 arcsec). Most galaxy SBPs are composed of a small compact bulge and an extended disk. 
The \textsc{CatSim} galaxy bulges have a steeper profile in the center when compared to the reference galaxy, while the disks are smaller on average (see \tabref{para_parameters}). 
The distribution of the difference between the disk and bulge magnitudes is shown in the bottom-right panel. Most of the galaxy flux is contained in the disk, with the disk being on average 2.5 magnitudes brighter than the bulge.

The galaxies selected here are not completely representative of the galaxies in the LSST lensing sample. 
However, the criteria are conservative in that they are not expected to reduce the estimated CG bias on LSST shear estimates. While the criteria could potentially introduce selection bias, the focus of this study is to isolate the impact of only color gradients.

\subsubsection{Results of CG shear bias measurements for \textsc{CatSim} galaxies}
\label{sec:cg_catsim}
\begin{figure}[!ht]
\centering\includegraphics[width=0.9\linewidth]{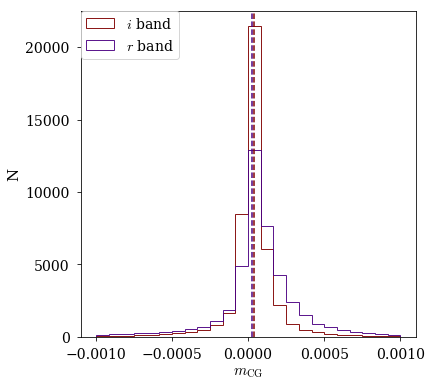}
\caption{Histograms of CG multiplicative shear bias in the $r$ and $i$ bands for \textsc{CatSim} galaxies. The almost completely overlapping dashed vertical lines denote the mean in each band.}
\label{fig:cat_para_ri}
\end{figure}

For each galaxy in the \textsc{CatSim} catalog we draw an equivalent galaxy with no CG and measure the shear response when a shear of $g=(0.01,0.01)$ is applied to both galaxies. 
A histogram of the estimated CG bias $m_{\rm CG}$ in the $r$ and $i$ bands is shown in \figref{cat_para_ri}, with their means shown as dashed vertical lines. 
The bias values are much smaller than LSST requirements. 
The mean and median CG biases are $(2.69 \pm 0.03) \times 10^{-5} $ and $6.17 \times 10^{-5}$ in $r$ band, and 
$(3.72 \pm 0.01) \times 10^{-5}$ and $2.58 \times 10^{-5}$ in $i$ band -- in each case, much smaller than LSST requirements.
 
\section{Estimating CG bias with chromatic real galaxies}
\label{sec:cg_real}
To place more robust limits on the expected CG shear bias for LSST we estimate CG bias for a large sample of realistic galaxies with color gradients. We use galaxy images from the Hubble Space Telescope (HST) to obtain realistic galaxy morphologies and then draw them as seen by LSST using the {\tt ChromaticRealGalaxy} module in {\tt GalSim}.

In a separate note\footnote{http://github.com/sowmyakth/measure\_cg\_bias/raw/master/\\
pdfs/Reducing\_AEGIS\_gal.pdf}, we describe the data reduction process used to produce postage-stamp images of 27k galaxies observed by the All-Wavelength Extended Groth strip International Survey (AEGIS) \citep{2007ApJ...660L...1D} in the HST V (F606W) and I (F814W) bands. Our goal is to use the V- and I-band images of galaxies in the AEGIS catalog to create their profiles in $r$ and $i$ LSST bands, while preserving their chromatic features. We can thus estimate the CG bias when these galaxies are imaged by LSST.

\subsection{{\tt ChromaticRealGalaxy} model}
\label{sec:crg}
{\tt ChromaticRealGalaxy}\footnote{http://github.com/GalSim-developers/GalSim/blob/master/\\
devel/modules/CGNotes.pdf}
models multi-band images of individual galaxies as chromatic PSF convolutions (and integrations over wavelength) with a sum of profiles separable into spatial and spectral components. This decomposition can be thought of as a constrained chromatic deconvolution of the multi-band images by the associated PSFs.

The pre-convolution chromatic surface brightness profile of the galaxy, $f_{\rm CG}(\vec{x}, \lambda)$, is modelled as a sum of two or more separable chromatic surface brightness profiles, each with a particular asserted SED:
\begin{equation}
f_{\rm CG}(\vec{x}, \lambda) = \sum_j a_j(\vec{x}) S_j(\lambda),
\label{eqn:crg_sbp}
\end{equation}
where $S_j(\lambda)$ is the $j$th SED asserted as part of the decomposition, and $a_j(\vec{x})$ is the spatial component of the
$j$th separable chromatic profile.

The observed image in the $i$th band is
\begin{eqnarray}
    I_i^{\rm obs}(\vec{x})
    & = & \int T_i(\lambda) \big[\Pi(\vec{x}, \lambda) \ast f(\vec{x}, \lambda)\big] \,  d\lambda + \eta_i(\vec{x}) \nonumber \\
    &=& \int T_i(\lambda) \sum_j S_j(\lambda) \left[\Pi(\vec{x}, \lambda) \ast a_j(\vec{x})\right] \,  d\lambda \nonumber \\
    & & + \,  \, \eta_i(\vec{x}),
\end{eqnarray}
where $\eta_i(\vec{x})$ corresponds to (potentially spatially correlated) Gaussian noise in the $i$th band image. The noise $\eta_i(\vec{x})$ is related to
the noise covariance function via $\langle\eta_i(\vec{x}_l) \eta_i(\vec{x}_m)\rangle =
\xi_i(\vec{x}_l - \vec{x}_m)$, where angle brackets indicate averaging over realizations of the
noise. 

The convolution is easier to work with in Fourier space where it becomes a mode-by-mode product. The model in Fourier space is
\begin{eqnarray}
    \tilde{I}_i^{\rm obs}(\vec{k})
    &=& \int T_i(\lambda) \sum_j S_j(\lambda) \tilde{\Pi}(\vec{k}, \lambda) \tilde{a}_j(\vec{k}) \,  d\lambda + \tilde{\eta}(\vec{k}) \nonumber \\
    &=& \sum_j \big[\int T_i(\lambda) S_j(\lambda) \tilde{\Pi}(\vec{k}, \lambda)  \, d\lambda\big] \tilde{a}_j(\vec{k}) + \tilde{\eta}(\vec{k}) \nonumber \\
    &=&  \sum_j \tilde{\Pi}^\mathrm{eff}_{ij}(\vec{k}) \tilde{a}_j(\vec{k}) + \tilde{\eta}(\vec{k}), 
\end{eqnarray}
where the effective PSF for the $i$th band and $j$th SED is
\begin{equation}
  \tilde{\Pi}^\mathrm{eff}_{ij}(\vec{k}) = \int T_i(\lambda) S_j(\lambda) \tilde{\Pi}(\vec{k}, \lambda)  \, d\lambda.
\end{equation}
{\tt ChromaticRealGalaxy} solves for the (complex-valued) $\tilde{a}_j(\vec{k})$ while propagating the statistics of the noise.

\subsection{Limitations of using real galaxy images}
\label{sec:limitations}
We use {\tt ChromaticRealGalaxy} (CRG) to model real HST galaxies in an LSST observing band and then convolve the images with the expected LSST PSF and add the sky noise expected at 10-year LSST depth. We assume here that the observing conditions for HST and LSST are known perfectly. For real images where the SED of the galaxies are unknown, we approximate the SEDs of the CRG components as polynomials. Thus, the accuracy with which the CRG algorithm reproduces chromatic features depends on how different the bandpasses of input and target surveys are, and the assumed model of the position-dependent SEDs. Since we use only two images of the HST galaxy (V and I band), the SEDs are assumed to be polynomials of order 0 and 1 (linear SED).

The simulations of parametric galaxies analyzed in \secref{cg_para} are noise free. Bias from CG is computed as the difference in the shear estimator measured from a galaxy with CG and the equivalent galaxy with no CG. The assumption was that biases from all sources other then color gradients would act equally on both measurements and thus cancel in the difference, giving the bias from CG only. However, each real galaxy image includes noise in the signal in each pixel\footnote{We assume that fluctuations in pixel values are dominated by Poisson fluctuations in a large number of detected electrons so that we can neglect the relatively small noise contributions expected from dark current and readout noise.} (pixel noise) while the image of the equivalent galaxy with no CG has different noise in each pixel.
This results in unequal noise bias in the shear measurements for the two galaxies. Therefore, the difference in shear measurements (\eqnref{cg}) no longer isolates bias from only CG; rather it is a combination of CG bias and residual noise bias.

Therefore, before using CRG to model real HST galaxies, we test CRG on simulated parametric galaxies where the truth is known. The detailed procedure used to test the impacts of the linear approximation for the SED and pixel noise while using CRG to model the reference galaxy and the \textsc{CatSim} galaxies is described in \appref{testing}. 
We summarize the results here.
\begin{enumerate}
    \item 
For reference galaxy with extreme color gradients the redshift-averaged error in CG shear bias caused by modelling the galaxy SEDs as linear is $\mathcal{O}(10^{-4})$; however, the linear SED model tends to smooth the variation in bias with redshift, leading to a total variation in bias over the redshift range $[0,1.2]$ that is a factor of $\approx$ 2 less for the linear SED compared to the true SED.
\item 
For simulated HST-like \textsc{CatSim} galaxies (modeled by CRG), 
the value of the CG bias estimate exhibits a strong dependence on SNR. 
 The measured bias diverges from the noise-free CG bias estimate by $\mathcal{O}(10^{-3})$ for HST I-band SNR $\lesssim$ 100. 
\item 
For simulated HST-like \textsc{CatSim} galaxies galaxies with I-band SNR $>$ 200, the mean bias is $m_{\rm CG} = (-2.53 \pm 0.27) \times{10^{-4}}$ in the $r$ band and $m_{\rm CG} = (0.47 \pm 0.10) \times{10^{-4}}$ in the $i$ band. The mean CG bias for the noise-free parametric simulations of the same sample of galaxies is $m_{\rm CG} = (-0.02 \pm 0.03) \times{10^{-4}}$ in the $r$ band and $m_{\rm CG} = (0.25 \pm 0.02) \times{10^{-4}}$ in the $i$ band.
\end{enumerate}

Thus, by using noisy images of \textsc{CatSim} galaxies, we find that our method of using an ``equivalent galaxy" to minimize the impact of effects other than color gradients does not completely eliminate the impact of ``pixel noise" for smaller values of SNR. Therefore, we expect the results from (noisy) AEGIS galaxies to be impacted by residual noise bias for small SNR. However, we do not observe a significant impact of pixel noise for high SNR galaxies in \textsc{CatSim} and therefore, importantly, we can assume that our method measures color gradient bias (with minimal contamination from residual noise bias) for high SNR AEGIS galaxies.

\subsection{Estimating CG bias with AEGIS galaxies}
\label{sec:aeg_survey}
\begin{figure}[ht!]
\centering\includegraphics[width=0.85\linewidth]{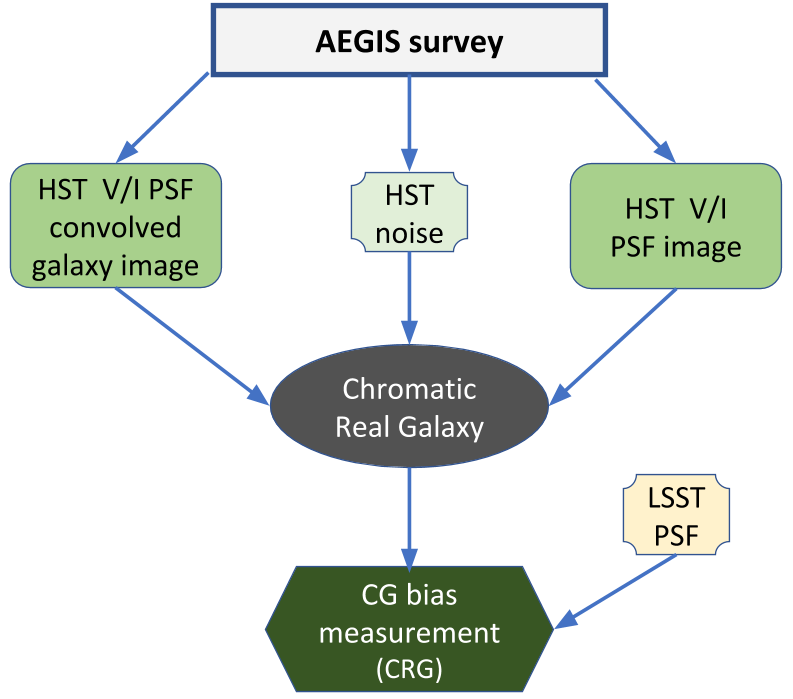}
\caption{Flow chart describing the method for measuring CG bias for real galaxies from the AEGIS survey.}
\label{fig:aeg_flow}
\end{figure}

Having demonstrated that {\tt ChromaticRealGalaxy} is able to reproduce color gradient bias results for galaxies with high SNR, we now apply our analysis to real galaxies with color gradients.
A flowchart depicting the methodology to estimate CG bias from real HST galaxies in the AEGIS survey is shown in \figref{aeg_flow}. The V- and I-band images of galaxies that pass certain selection criteria are input to {\tt ChromaticRealGalaxy} to model the galaxy SBP. Along with the PSF-convolved images, CRG also requires as input the noise correlation functions and PSF images in the V and I bands. The modelled chromatic SBP of each galaxy is then used to estimate CG bias when seen by LSST as described in \secref{psf}.

\subsubsection{Creating the AEGIS catalog}
The AEGIS imaging data is composed of a mosaic pattern of $21 \times 3 = 63$ contiguous HST ``tiles" covering an effective area of $∼10.1'\times 70.5' = 710.9$ arcmin$^2$, with a four pointing dither pattern for each tile \citep{2007ApJ...660L...1D}. The dithered observations for each pointing and band were combined (``drizzled'') using {\tt MultiDrizzle} to produce coadded images with a pixel scale of 0.03 arcsec \citep{2007ApJS..172..203R}. 

The procedure used to create the catalog, based on \cite{bemi}, is similar to that used for the weak lensing catalog for the HST ACS COSMOS survey~\citep{2007ApJS..172..219L}.
We use version 2.8.6 of the 
{\tt  SExtractor}\footnote{https://www.astromatic.net/software/sextractor} \citep{1996A&AS..117..393B} package to produce a source catalog of positions and various photometric parameters. Detection was performed on a coaddition image of both bands, and photometric measurements then performed on each band at the previously detected locations. Unreliable regions such as tile boundaries, diffraction spikes and ``ghosts'' due to internal reflections are masked.
CRG modelling requires postage stamp images of individual galaxies. Thus if flux from a neighboring object lies within a stamp, it is replaced with noise. If the overlap with the neighbor is too significant, then the galaxy is not selected for analysis. 

The final catalog consists of 26,517 galaxies that satisfy the following criteria:
\begin{itemize}
\item The object was detected in both V and I bands.
\item The object is not classified as a star in either band.
\item The object is not in a masked region in either band.
\item The object has magnitude brighter than $25.2$ in the I band. 
\item The postage stamp does not contain flux from a neighboring object (no significant overlap).
\end{itemize}
For 24,635 of these galaxies, shapes were successfully measured in the ring test for the galaxy and for its equivalent with no CG.

We model the PSF by comparing stars in a given HST tile to star fields drawn from {\tt Tiny Tim} ray-tracing software \citep{1993ASPC...52..536K}.
The variations in the size and ellipticity of the PSF across the focal plane of the HST ACS is dominated by the effective focus, which changes due to thermal expansion and contraction of the HST. {\tt Tiny Tim} simulates the variation of the PSF across the field for different focal lengths. The PSF variation across an image is characterized by comparing the measured PSF ellipticity for stars in a field to the {\tt Tiny Tim} predictions for different focus offsets, and finding the focus that best matches the measured and predicted PSFs across the field \citep{2007ApJS..172..219L}.
The PSF for each galaxy is taken to be the {\tt Tiny Tim} model PSF image whose location in the {\tt Tiny Tim} grid is closest to that galaxy's position in the focal plane, for the best-fit focus offset. 

The correlated noise in the two bands is estimated from empty regions in the AEGIS fields.
A detailed description of how individual galaxy images, their corresponding PSFs and the noise correlation functions are determined can be found in the note described in \secref{cg_real}. We also include in our final catalog the spectroscopic redshifts for 3763 matched galaxies from the DEEP2 galaxy redshift survey, Data Release 4 \citep{2013ApJS..208....5N}.

\subsubsection{Characteristics of AEGIS catalog}
\label{sec:aeg_charac}
\begin{figure*}[!ht]
\centering\includegraphics[width=0.9\linewidth]{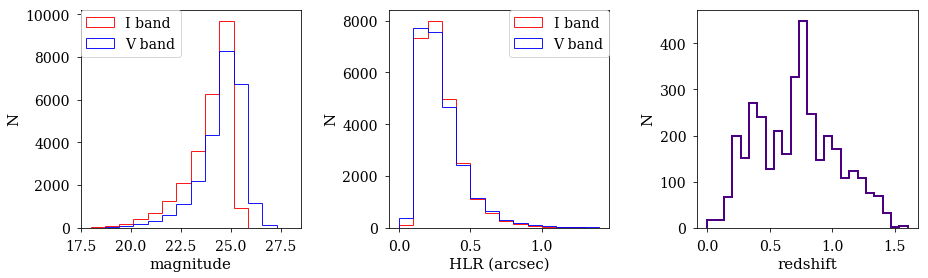}
\caption{Distributions of AEGIS galaxy parameters. 
Left: V- and I-band magnitude. A selection criterion of I$<25.2$ was applied in the generation of the catalog.
Center: Half-light radius in arcseconds. 
Right: Spectroscopic redshift for 14\% of galaxies in the survey.}
\label{fig:aeg_gal}
\end{figure*}

Characteristics of the selected AEGIS galaxies are shown in \figref{aeg_gal}. The left panel shows the distribution of magnitudes in V and I bands. The selection criteria ensure that all galaxies have I $<$ 25.2. The center panel shows the distribution of galaxy HLR in arcseconds; 32\% of the galaxies have HLR less than or equal to the LSST pixel scale of 0.2 arcsec. 
The magnitude and HLR were estimated with {\tt SExtractor}. The right panel shows the redshift distribution for the 14\% of the galaxies for which spectroscopic redshift estimates are available in the DEEP2 Galaxy Redshift Survey.

\begin{figure}[!ht]
\centering\includegraphics[width=1.\linewidth]{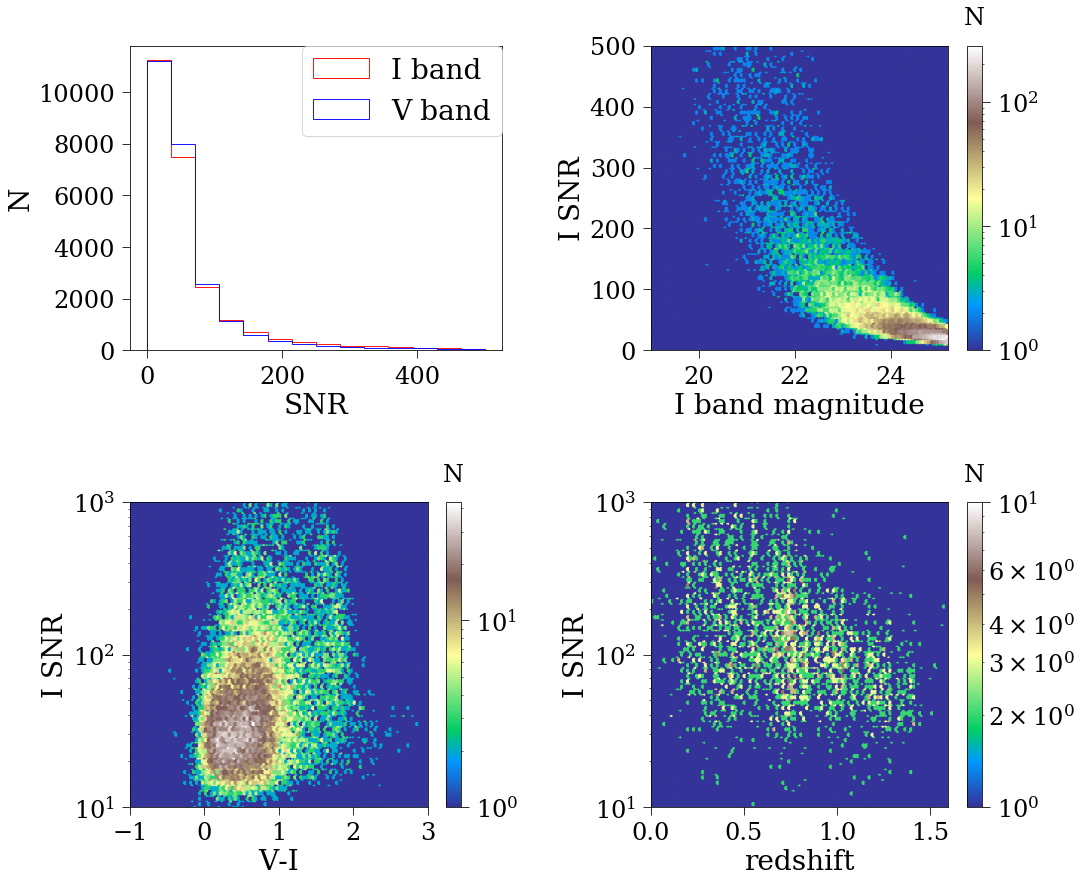}
\caption{HST SNR of AEGIS galaxies. 
Top left: SNR for HST V and I bands.
Top right: I-band SNR versus I-band magnitude.
Bottom left: I-band SNR versus V$-$I color (left). Bottom right: I-band SNR versus redshift for the 14\% of galaxies for which spectroscopic redshift estimates are available.}
\label{fig:aeg_gal_snr}
\end{figure}

Since CG bias estimates show a large dependence on galaxy SNR, we show the distribution of galaxy SNR in HST V and I bands in \figref{aeg_gal_snr} (top left panel); the mean SNRs are 76 and 88 in V and I bands, respectively. 
The top-right panel in \figref{aeg_gal_snr} shows the correlation of I-band SNR with magnitude. 
A selection criterion of HST I-band SNR $>$ 200 excludes galaxies with I-band magnitude fainter than 23.2.
In the bottom panels, we plot SNR in I band as a function of V$-$I color (left), and redshift (right). 
The selection cut of I-band SNR $>$ 200 eliminates 97\% of galaxies with $z>1.2$ from the sample.

\subsubsection{Results of CG bias analysis of AEGIS galaxies}
\begin{figure*}[!ht]
\centering\includegraphics[width=1\linewidth]{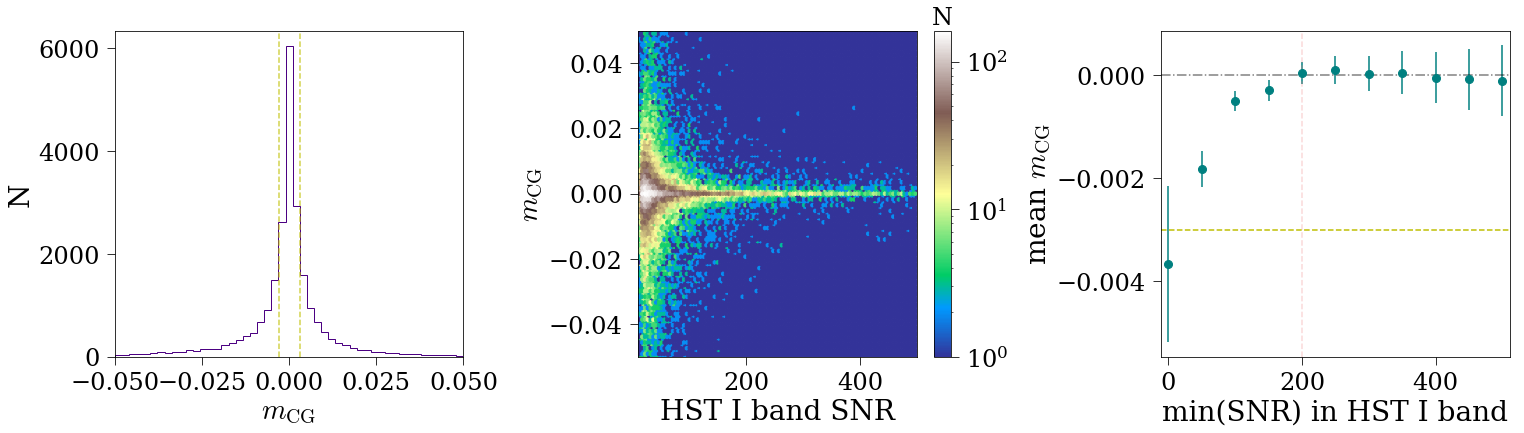}
\caption{Left: Histogram of estimated multiplicative shear bias due to CG, $m_{\rm CG}$, for $r$-band images computed with {\tt ChromaticRealGalaxy} from all AEGIS galaxies. 
Center: $m_{\rm CG}$ as a function of HST I-band SNR.
Right: Mean of estimated CG bias for galaxies for which SNR is greater than min(SNR).
At low SNR, the magnitude of the estimated CG bias is dominated by noise bias. 
The yellow dashed line shows the LSST DESC requirement of 0.003 on the total systematic uncertainty in the redshift-dependent shear calibration.
}
\label{fig:aeg_mCG_results}
\end{figure*}

The results of the CG bias analysis of AEGIS galaxies when observed in the LSST $r$ band are shown in \figref{aeg_mCG_results}. The plot on the left shows the distribution of estimated bias. The mean multiplicative shear bias is $m_{\rm CG} = (-3.7 \pm 1.6) \times{10^{-3}}$ 
for all galaxies in the sample. However, as shown in the center panel, the large spread in the estimated bias is due to the low SNR galaxies. 
In the right panel, we show the mean value of $m_{\rm CG}$ for galaxies with I-band SNR greater than a minimum value ranging from 0 to 500. 
The dashed black horizontal line corresponds to zero bias. As we exclude low SNR galaxies from the sample, the measured CG bias values approach zero. The error bars correspond to the statistical uncertainties on the mean.
The mean CG bias for the 1900 galaxies with I-band SNR $>$ 200 is $m_{\rm CG} = (0.4 \pm 2.2) \times{10^{-4}}$.

\begin{figure*}[!ht]
\centering\includegraphics[width=0.8\linewidth]{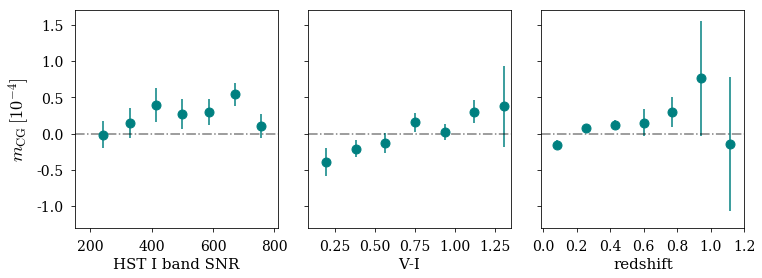}
\caption{Mean CG bias estimates for $r$-band images generated from AEGIS galaxies with I-band SNR $>200$ as a function of HST I-band SNR (left) and V$-$I color (center) for all galaxies, and as a function of redshift (right). The dependence on SNR and color is shown for 1900 galaxies while the redshift dependence is shown for the 1043 galaxies for which spectroscopic redshift estimates are available.}
\label{fig:aeg_mCG_bin}
\end{figure*}

In \figref{aeg_mCG_bin}, we show the dependence of the CG bias estimate $m_{\rm CG}$ on SNR, V$-$I color, and redshift for galaxies with I-band SNR $>200$. The SNR and color dependence is depicted for the 1900 galaxies with I-band SNR $>$ 200, while the redshift dependence is plotted for 1043 of these galaxies for which spectroscopic redshift estimates are available.
The dots denote the mean value for each bin while the error bars denote the statistical uncertainty on the mean. The CG bias values do not show a statistically significant dependence on galaxy color.
The bias values show a small dependence on SNR; however, the magnitudes of the mean biases are lower than the LSST requirement. 

\begin{figure*}[!ht]
\centering\includegraphics[width=.9\linewidth]{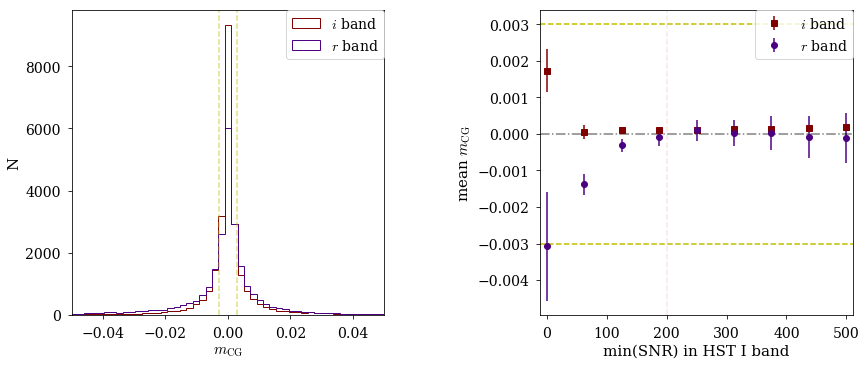}
\caption{Left: Histogram of multiplicative shear bias from CG, $m_{\rm CG}$, in the $i$ band (red) and $r$ band (blue) for AEGIS galaxies. Right: Mean $m_{\rm CG}$ in the $i$ band (red) and $r$ band (blue) for galaxies with I-band SNR $>$ min(SNR). The yellow dashed lines show the LSST DESC requirement of 0.003 on the total systematic uncertainty in the redshift-dependent shear calibration.}
\label{fig:aeg_mCG_i_r}
\end{figure*}

We compare CG bias estimates for the AEGIS galaxies when observed in LSST $i$ and $r$ bands in \figref{aeg_mCG_i_r}. We show bias estimates for 24,635 galaxies from the AEGIS sample for which bias estimates were successfully computed in \emph{both} bands.  The left panel shows the distribution of bias estimates. The $i$-band CG bias estimate distribution has smaller tails than the $r$ band, with a mean $m_{\rm CG} = (1.7 \pm 0.5) \times{10^{-3}}$ for the entire sample. The right panel plots the mean CG bias for galaxies above a minimum I-band SNR. The mean bias quickly converges to zero, once the low SNR galaxies are excluded from the sample. 
For galaxies with I-band SNR $>$ 200, the mean multiplicative shear bias in the $i$ band is $m_{\rm CG} = (7.8 \pm 3.6) \times{10^{-5}}$.

\section{Limitations of this study}
\label{sec:limitations_study}
We summarize here the limitations of this analysis and how it could be extended to future studies.
\begin{enumerate}
    \item Our method for isolating CG bias breaks down when galaxy SBPs are modelled from noisy images. This limits our analysis of real images to high SNR galaxies.
    Due to this limitation, we are only able to measure CG bias for real galaxies with redshift $<$ 1.2.
    Estimating CG bias for higher redshift galaxies will require a statistically significant sample of high SNR real images or high fidelity simulations of noise-free galaxies with realistic morphologies.
    \item Our methodology assumes a linear response to the applied shear, which is true for weak lensing measurements. However, for lensing by galaxy clusters and in the strong lensing regime when this approximation no longer holds, color gradient bias could result in non-negligible biases in shear estimates.
    \item We studied one source of PSF chromaticity: wavelength-dependent PSF size. However, several chromatic effects, including differential chromatic refraction, refraction in optics, and sensor effects, can also produce a wavelength-dependent PSF ellipticity.
    \item The fidelity of color gradient models in the samples of galaxies used in this study is limited.
    In {\tt ChromaticRealGalaxy}, the SEDs for the AEGIS galaxies, as would be seen by LSST, are modelled as simple linear functions of wavelength because only two HST bands are available. 
   The parametric galaxies in the \textsc{CatSim} sample are described as simple Sersic bulge + disk profiles. 
   As our knowledge of color gradients as a function of redshift improves, future studies could use more realistic CG models, including active galactic nuclei, star forming regions, dust lanes, etc.
\end{enumerate}

\section{Conclusions}
We extend the method employed in \cite{2013MNRAS.432.2385S} for Euclid-type observations to isolate the effect of color gradients on weak lensing shear measurements for observations in the LSST $r$ and $i$ bands. The bias originates when shapes of galaxies with color gradients, observed with a chromatic PSF, are measured with some form of spatial weighting. 
The bias depends only weakly on the size of the PSF, but depends linearly on the magnitude of the exponent $\alpha$ in its wavelength dependence $\lambda^\alpha$ (\figref{ref_gal_psf_cg}). The bias is very sensitive to the size of the weight function (\figref{ref_gal_weight}). 
We estimate CG shear bias for three samples as observed with LSST: reference parametric galaxies with large color gradients, an ensemble of parametric galaxies with more realistic color gradients, and galaxies with realistic morphology. The results are summarized in \tabref{summary}.

For noise-free parametric galaxy simulations, 
the value of half the maximum span of $m_{\rm CG}(z)$ in the redshift range [0, 1.2] is $\leq 1.5 \times 10^{-3}$ for the reference galaxies with extreme color gradients and $\leq 10^{-4}$ for \textsc{CatSim} galaxies. For input AEGIS galaxies with pixel noise, the estimated bias shows a strong dependence on SNR due to contributions from effects other than CG.
However, for AEGIS galaxies with HST I-band SNR $> 200$, the magnitude of the mean estimated bias in the LSST $r$ band is $(0.4 \pm 2.2)\times{10^{-4}}$, while the value of half the maximum span of $m_{\rm CG}(z)$ in the redshift range [0, 1.2] is $\leq 1.5 \times 10^{-4}$. 
Therefore, for both the noise-free parametric galaxies and for the AEGIS galaxies with SNR $> 200$, the half-maximum span is less than the LSST full-depth requirement of $0.003$ on the total systematic uncertainty in the redshift-dependent shear calibration $m_z$ (\eqnref{tomo-z-equation}).
This result is important because even self-calibrating shear estimation techniques (e.g., {\tt MetaCal} \citep{2017arXiv170202600H}) assume accurate knowledge of the PSF, and the effects of color gradients on the per-galaxy PSF could be very challenging to predict and correct. 
We are thus optimistic that color gradients will not be a source of limiting systematic uncertainty for LSST; however, the limitations listed in \secref{limitations_study} should be investigated as more observations and studies of color gradients in galaxies become available.

\begin{deluxetable*}{c|ccc}
\tablecaption{Results for multiplicative shear bias statistics for $m_{\rm CG}(z)$, due to color gradients. \label{tab:summary}}
\tablehead{
\colhead{Sample} &\colhead{\makecell{Half maximum span of $m_{\rm CG}(z)$\tablenotemark{a}
\\ in $r$ band}} & \colhead{\makecell{Mean $m_{\rm CG}$ \\ in $r$ band}} & \colhead{\makecell{Mean $m_{\rm CG}$ \\ in $i$ band}}}
\startdata
\makecell{Reference galaxy with \\extreme color gradients (parametric)}\tablenotemark{b} & $1.5 \times 10^{-3}$ (see \figref{ref_gal_diff_sed}) &
$(8.7 \pm 0.6) \times 10^{-4}$ & -- \\
\textsc{CatSim} galaxies (parametric)\tablenotemark{b} & $\leq 10^{-4}$ (see \figref{cat_mCG_bin}) & $(0.268 \pm 0.003) \times{10^{-4}}$ & $(0.372 \pm 0.001) \times{10^{-4}}$ \\
AEGIS galaxies (CRG)\tablenotemark{c}& $\leq 1.5 \times 10^{-4}$(see \figref{aeg_mCG_bin}) & $(0.4 \pm 2.2) \times{10^{-4}}$ &
$(0.8 \pm 0.4) \times{10^{-4}}$ \\
\enddata
\tablenotetext{a}{Half maximum span of $m_{\rm CG}$ over redshift range [0, 1.2]; see \secref{LSST_requirements} for motivation for requirement of 0.003 on this value.}
\tablenotetext{b}{Values computed for noise-free parametric simulations.}
\tablenotetext{c}{Values computed for chromatic real galaxies (CRG) generated from images with HST I-band SNR\,$>$\,200.}
\end{deluxetable*}

\subsection*{Acknowledgments}
This paper has undergone internal review in the LSST Dark Energy Science Collaboration. The internal reviewers were Chihway Chang, Douglas Clowe, and Mike Jarvis.

J.E.M.\  and P.R.B.\  provided motivation for the study and guidance throughout the analysis. 
S.K.\  produced the simulated images, prepared the AEGIS-based images, and performed the analysis. 
J.E.M.\  wrote the {\tt ChromaticRealGalaxy} module in {\tt GalSim}, to simulate LSST galaxies from HST images, and offered technical advice. 
All authors discussed interpretation of the results. 
S.K.\  wrote the initial draft of the manuscript;   P.R.B., S.K., and J.E.M.\  all contributed to the final draft.

We thank the developers of the {\tt GalSim} galaxy simulation package and Rachel Mandelbaum and Michael Jarvis in particular, for advice and feedback on incorporating AEGIS galaxies  in {\tt GalSim}. We thank Bradley Emi, Andr\'es Plazas, Anton Koekemoer, and Jason Rhodes for providing the AEGIS dataset and for help in data reduction. 

The LSST DESC acknowledges ongoing support from the Institut National de Physique Nucl\'eaire et de Physique des Particules in France; the Science \& Technology Facilities Council in the United Kingdom; and the Department of Energy, the National Science Foundation, and the LSST Corporation in the United States.  
The LSST DESC uses resources of the IN2P3 Computing Center (CC-IN2P3--Lyon/Villeurbanne - France) funded by the Centre National de la Recherche Scientifique; the National Energy Research Scientific Computing Center, a DOE Office of Science User Facility supported by the Office of Science of the U.S.\ Department of Energy under Contract No.\ DE-AC02-05CH11231; STFC DiRAC HPC Facilities, funded by UK BIS National E-infrastructure capital grants; and the UK particle physics grid, supported by the GridPP Collaboration.  This work was performed in part under DOE Contract DE-AC02-76SF00515 (SLAC), DOE Grant DE-SC0019351 (Stanford), and NSF Grant PHY-1404070 (Stanford).

\software{{\tt GalSim} \citep{2015A&C....10..121R}, {\tt SExtractor} \citep{1996A&AS..117..393B}, 
Astropy \citep{2013A&A...558A..33A, 2018AJ....156..123A}, 
NumPy \citep{2011CSE....13b..22V}, 
matplotlib \citep{2007CSE.....9...90H}.}
\bibliography{lsstdesc,main}

\begin{thebibliography}{}
\expandafter\ifx\csname natexlab\endcsname\relax\def\natexlab#1{#1}\fi

\bibitem[{{Albrecht} {et~al.}(2006){Albrecht}, {Bernstein}, {Cahn}, {Freedman},
  {Hewitt}, {Hu}, {Huth}, {Kamionkowski}, {Kolb}, \&
  {Knox}}]{2006astro.ph..9591A}
{Albrecht}, A., {Bernstein}, G., {Cahn}, R., {et~al.} 2006, arXiv e-prints,
  astro

\bibitem[{{Amiaux} {et~al.}(2012){Amiaux}, {Scaramella}, {Mellier}, {Altieri},
  {Burigana}, {Da Silva}, {Gomez}, {Hoar}, {Laureijs}, \&
  {Maiorano}}]{2012SPIE.8442E..0ZA}
{Amiaux}, J., {Scaramella}, R., {Mellier}, Y., {et~al.} 2012, in Society of
  Photo-Optical Instrumentation Engineers (SPIE) Conference Series, Vol. 8442,
  \procspie, 84420Z

\bibitem[{{Astropy Collaboration} {et~al.}(2013){Astropy Collaboration},
  {Robitaille}, {Tollerud}, {Greenfield}, {Droettboom}, {Bray}, {Aldcroft},
  {Davis}, {Ginsburg}, \& {Price-Whelan}}]{2013A&A...558A..33A}
{Astropy Collaboration}, {Robitaille}, T.~P., {Tollerud}, E.~J., {et~al.} 2013,
  \aap, 558, A33

\bibitem[{{Astropy Collaboration} {et~al.}(2018){Astropy Collaboration},
  {Price-Whelan}, {Sip{\H{o}}cz}, {G{\"u}nther}, {Lim}, {Crawford}, {Conseil},
  {Shupe}, {Craig}, \& {Dencheva}}]{2018AJ....156..123A}
{Astropy Collaboration}, {Price-Whelan}, A.~M., {Sip{\H{o}}cz}, B.~M., {et~al.}
  2018, \aj, 156, 123

\bibitem[{{Bartelmann} \& {Schneider}(2001)}]{2001PhR...340..291B}
{Bartelmann}, M., \& {Schneider}, P. 2001, \physrep, 340, 291

\bibitem[{{Bernstein} \& {Jarvis}(2002)}]{2002AJ....123..583B}
{Bernstein}, G.~M., \& {Jarvis}, M. 2002, \aj, 123, 583

\bibitem[{{Bertin} \& {Arnouts}(1996)}]{1996A&AS..117..393B}
{Bertin}, E., \& {Arnouts}, S. 1996, \aaps, 117, 393

\bibitem[{{Coleman} {et~al.}(1980){Coleman}, {Wu}, \&
  {Weedman}}]{1980ApJS...43..393C}
{Coleman}, G.~D., {Wu}, C.-C., \& {Weedman}, D.~W. 1980, \apjs, 43, 393

\bibitem[{{Connolly} {et~al.}(2014){Connolly}, {Angeli}, {Chandrasekharan},
  {Claver}, {Cook}, {Ivezic}, {Jones}, {Krughoff}, {Peng}, {Peterson}, {Petry},
  {Rasmussen}, {Ridgway}, {Saha}, {Sembroski}, {vanderPlas}, \&
  {Yoachim}}]{2014SPIE.9150E..14C}
{Connolly}, A.~J., {Angeli}, G.~Z., {Chandrasekharan}, S., {et~al.} 2014, in
  Society of Photo-Optical Instrumentation Engineers (SPIE) Conference Series,
  Vol. 9150, Society of Photo-Optical Instrumentation Engineers (SPIE)
  Conference Series, 14

\bibitem[{{Cypriano} {et~al.}(2010){Cypriano}, {Amara}, {Voigt}, {Bridle},
  {Abdalla}, {R{\'e}fr{\'e}gier}, {Seiffert}, \&
  {Rhodes}}]{2010MNRAS.405..494C}
{Cypriano}, E.~S., {Amara}, A., {Voigt}, L.~M., {et~al.} 2010, \mnras, 405, 494

\bibitem[{{Davis} {et~al.}(2007){Davis}, {Guhathakurta}, {Konidaris}, {Newman},
  {Ashby}, {Biggs}, {Barmby}, {Bundy}, {Chapman}, {Coil}, {Conselice},
  {Cooper}, {Croton}, {Eisenhardt}, {Ellis}, {Faber}, {Fang}, {Fazio},
  {Georgakakis}, {Gerke}, {Goss}, {Gwyn}, {Harker}, {Hopkins}, {Huang},
  {Ivison}, {Kassin}, {Kirby}, {Koekemoer}, {Koo}, {Laird}, {Le Floc'h}, {Lin},
  {Lotz}, {Marshall}, {Martin}, {Metevier}, {Moustakas}, {Nandra}, {Noeske},
  {Papovich}, {Phillips}, {Rich}, {Rieke}, {Rigopoulou}, {Salim},
  {Schiminovich}, {Simard}, {Smail}, {Small}, {Weiner}, {Willmer}, {Willner},
  {Wilson}, {Wright}, \& {Yan}}]{2007ApJ...660L...1D}
{Davis}, M., {Guhathakurta}, P., {Konidaris}, N.~P., {et~al.} 2007, \apjl, 660,
  L1

\bibitem[{{De Lucia} {et~al.}(2006){De Lucia}, {Springel}, {White}, {Croton},
  \& {Kauffmann}}]{2006MNRAS.366..499D}
{De Lucia}, G., {Springel}, V., {White}, S.~D.~M., {Croton}, D., \&
  {Kauffmann}, G. 2006, \mnras, 366, 499

\bibitem[{{Emi}(2018)}]{bemi}
{Emi}, B. 2018, Stanford Undergraduate Research Journal, 105

\bibitem[{{Er} {et~al.}(2018){Er}, {Hoekstra}, {Schrabback}, {Cardone},
  {Scaramella}, {Maoli}, {Vicinanza}, {Gillis}, \&
  {Rhodes}}]{2018MNRAS.476.5645E}
{Er}, X., {Hoekstra}, H., {Schrabback}, T., {et~al.} 2018, \mnras, 476, 5645

\bibitem[{{Fried}(1966)}]{1966JOSA...56.1372F}
{Fried}, D.~L. 1966, Journal of the Optical Society of America (1917-1983), 56,
  1372

\bibitem[{{Hirata} \& {Seljak}(2003)}]{2003MNRAS.343..459H}
{Hirata}, C., \& {Seljak}, U. 2003, \mnras, 343, 459

\bibitem[{{Huff} \& {Mandelbaum}(2017)}]{2017arXiv170202600H}
{Huff}, E., \& {Mandelbaum}, R. 2017, arXiv e-prints, arXiv:1702.02600

\bibitem[{{Hunter}(2007)}]{2007CSE.....9...90H}
{Hunter}, J.~D. 2007, Computing in Science and Engineering, 9, 90

\bibitem[{{Ivezic} {et~al.}(2008){Ivezic}, {Tyson}, {et~al.}}]{Overview}
{Ivezic}, Z., {Tyson}, J.~A., {et~al.} 2008, ArXiv e-prints, arXiv:0805.2366

\bibitem[{{Kaiser} {et~al.}(1995){Kaiser}, {Squires}, \&
  {Broadhurst}}]{1995ApJ...449..460K}
{Kaiser}, N., {Squires}, G., \& {Broadhurst}, T. 1995, \apj, 449, 460

\bibitem[{{Kennedy} {et~al.}(2016){Kennedy}, {Bamford}, {H{\"a}u{\ss}ler},
  {Baldry}, {Bremer}, {Brough}, {Brown}, {Driver}, {Duncan}, \&
  {Graham}}]{2016MNRAS.460.3458K}
{Kennedy}, R., {Bamford}, S.~P., {H{\"a}u{\ss}ler}, B., {et~al.} 2016, \mnras,
  460, 3458

\bibitem[{{Krist}(1993)}]{1993ASPC...52..536K}
{Krist}, J. 1993, in Astronomical Society of the Pacific Conference Series,
  Vol.~52, Astronomical Data Analysis Software and Systems II, ed. R.~J.
  {Hanisch}, R.~J.~V. {Brissenden}, \& J.~{Barnes}, 536

\bibitem[{{La Barbera} {et~al.}(2010){La Barbera}, {De Carvalho}, {De La Rosa},
  {Gal}, {Swindle}, \& {Lopes}}]{2010AJ....140.1528L}
{La Barbera}, F., {De Carvalho}, R.~R., {De La Rosa}, I.~G., {et~al.} 2010,
  \aj, 140, 1528

\bibitem[{{Leauthaud} {et~al.}(2007){Leauthaud}, {Massey}, {Kneib}, {Rhodes},
  {Johnston}, {Capak}, {Heymans}, {Ellis}, {Koekemoer}, \& {Le
  F{\`e}vre}}]{2007ApJS..172..219L}
{Leauthaud}, A., {Massey}, R., {Kneib}, J.-P., {et~al.} 2007, \apjs, 172, 219

\bibitem[{{LSST DESC} {et~al.}(2018){LSST DESC}, {Mandelbaum}, {Eifler},
  {Hlo{\v z}ek}, {Collett}, {Gawiser}, {Scolnic}, {Alonso}, {Awan}, {Biswas},
  {Blazek}, {Burchat}, {Chisari}, {Dell'Antonio}, {Digel}, {Frieman},
  {Goldstein}, {Hook}, {Ivezi{\'c}}, {Kahn}, {Kamath}, {Kirkby}, {Kitching},
  {Krause}, {Leget}, {Marshall}, {Meyers}, {Miyatake}, {Newman}, {Nichol},
  {Rykoff}, {Sanchez}, {Slosar}, {Sullivan}, \& {Troxel}}]{v1DESC-SRD}
{LSST DESC}, {Mandelbaum}, R., {Eifler}, T., {et~al.} 2018, ArXiv e-prints,
  arXiv:1809.01669v1

\bibitem[{{LSST Science Collaboration}(2009)}]{ScienceBook}
{LSST Science Collaboration}. 2009, ArXiv e-prints, arXiv:0912.0201

\bibitem[{{Luppino} \& {Kaiser}(1997)}]{1997ApJ...475...20L}
{Luppino}, G.~A., \& {Kaiser}, N. 1997, \apj, 475, 20

\bibitem[{{Mandelbaum} {et~al.}(2014){Mandelbaum}, {Rowe}, {Bosch}, {Chang},
  {Courbin}, {Gill}, {Jarvis}, {Kannawadi}, {Kacprzak}, {Lackner}, {Leauthaud},
  {Miyatake}, {Nakajima}, {Rhodes}, {Simet}, {Zuntz}, {Armstrong}, {Bridle},
  {Coupon}, {Dietrich}, {Gentile}, {Heymans}, {Jurling}, {Kent}, {Kirkby},
  {Margala}, {Massey}, {Melchior}, {Peterson}, {Roodman}, \&
  {Schrabback}}]{2014ApJS..212....5M}
{Mandelbaum}, R., {Rowe}, B., {Bosch}, J., {et~al.} 2014, \apjs, 212, 5

\bibitem[{{Mandelbaum} {et~al.}(2018){Mandelbaum}, {Lanusse}, {Leauthaud},
  {Armstrong}, {Simet}, {Miyatake}, {Meyers}, {Bosch}, {Murata}, {Miyazaki}, \&
  {Tanaka}}]{2018MNRAS.481.3170M}
{Mandelbaum}, R., {Lanusse}, F., {Leauthaud}, A., {et~al.} 2018, \mnras, 481,
  3170

\bibitem[{{Melchior} {et~al.}(2011){Melchior}, {Viola}, {Sch{\"a}fer}, \&
  {Bartelmann}}]{2011MNRAS.412.1552M}
{Melchior}, P., {Viola}, M., {Sch{\"a}fer}, B.~M., \& {Bartelmann}, M. 2011,
  \mnras, 412, 1552

\bibitem[{{Meyers} \& {Burchat}(2015)}]{2015ApJ...807..182M}
{Meyers}, J.~E., \& {Burchat}, P.~R. 2015, \apj, 807, 182

\bibitem[{{Nakajima} \& {Bernstein}(2007)}]{2007AJ....133.1763N}
{Nakajima}, R., \& {Bernstein}, G. 2007, \aj, 133, 1763

\bibitem[{{Newman} {et~al.}(2013){Newman}, {Cooper}, {Davis}, {Faber}, {Coil},
  {Guhathakurta}, {Koo}, {Phillips}, {Conroy}, \&
  {Dutton}}]{2013ApJS..208....5N}
{Newman}, J.~A., {Cooper}, M.~C., {Davis}, M., {et~al.} 2013, \apjs, 208, 5

\bibitem[{{Refregier} {et~al.}(2012){Refregier}, {Kacprzak}, {Amara}, {Bridle},
  \& {Rowe}}]{2012MNRAS.425.1951R}
{Refregier}, A., {Kacprzak}, T., {Amara}, A., {Bridle}, S., \& {Rowe}, B. 2012,
  \mnras, 425, 1951

\bibitem[{{Rhodes} {et~al.}(2007){Rhodes}, {Massey}, {Albert}, {Collins},
  {Ellis}, {Heymans}, {Gardner}, {Kneib}, {Koekemoer}, \&
  {Leauthaud}}]{2007ApJS..172..203R}
{Rhodes}, J.~D., {Massey}, R.~J., {Albert}, J., {et~al.} 2007, \apjs, 172, 203

\bibitem[{{Rowe} {et~al.}(2015){Rowe}, {Jarvis}, {Mandelbaum}, {Bernstein},
  {Bosch}, {Simet}, {Meyers}, {Kacprzak}, {Nakajima}, {Zuntz}, {Miyatake},
  {Dietrich}, {Armstrong}, {Melchior}, \& {Gill}}]{2015A&C....10..121R}
{Rowe}, B.~T.~P., {Jarvis}, M., {Mandelbaum}, R., {et~al.} 2015, Astronomy and
  Computing, 10, 121

\bibitem[{{Seitz} \& {Schneider}(1997)}]{1997A&A...318..687S}
{Seitz}, C., \& {Schneider}, P. 1997, \aap, 318, 687

\bibitem[{{Semboloni} {et~al.}(2013){Semboloni}, {Hoekstra}, {Huang},
  {Cardone}, {Cropper}, {Joachimi}, {Kitching}, {Kuijken}, {Lombardi}, \&
  {Maoli}}]{2013MNRAS.432.2385S}
{Semboloni}, E., {Hoekstra}, H., {Huang}, Z., {et~al.} 2013, \mnras, 432, 2385

\bibitem[{{S{\'e}rsic}(1963)}]{1963BAAA....6...41S}
{S{\'e}rsic}, J.~L. 1963, Boletin de la Asociacion Argentina de Astronomia La
  Plata Argentina, 6, 41

\bibitem[{{Springel} {et~al.}(2005){Springel}, {White}, {Jenkins}, {Frenk},
  {Yoshida}, {Gao}, {Navarro}, {Thacker}, {Croton}, {Helly}, {Peacock}, {Cole},
  {Thomas}, {Couchman}, {Evrard}, {Colberg}, \& {Pearce}}]{2005Natur.435..629S}
{Springel}, V., {White}, S.~D.~M., {Jenkins}, A., {et~al.} 2005, \nat, 435, 629

\bibitem[{{van der Walt} {et~al.}(2011){van der Walt}, {Colbert}, \&
  {Varoquaux}}]{2011CSE....13b..22V}
{van der Walt}, S., {Colbert}, S.~C., \& {Varoquaux}, G. 2011, Computing in
  Science and Engineering, 13, 22

\bibitem[{{Voigt} {et~al.}(2012){Voigt}, {Bridle}, {Amara}, {Cropper},
  {Kitching}, {Massey}, {Rhodes}, \& {Schrabback}}]{2012MNRAS.421.1385V}
{Voigt}, L.~M., {Bridle}, S.~L., {Amara}, A., {et~al.} 2012, \mnras, 421, 1385

\end{thebibliography}

\begin{appendix}
\section{Testing {\tt{ChromaticRealGalaxy}}}
\label{app:testing}
In this section we describe the tests conducted to estimate the accuracy of the {\tt ChromaticRealGalaxy} (CRG) algorithm for reproducing chromatic features from noisy images. 
We simulate parametric galaxies as they would be observed by HST (henceforth called ``HST-like'' galaxies) -- \ie, convolved with an HST-like chromatic PSF and observed in the HST V and I bands with correlated pixel noise. {\tt ChromaticRealGalaxy} is then used to model the SBP of the HST-like galaxies in order to estimate the CG bias when observed by LSST. We then compare this estimated CG bias to the CG bias estimated in \secref{cg_para} from noise-free parametric galaxies simulated directly in LSST bands.

\subsection{Testing CRG with the reference galaxy with extreme color gradients}
\label{app:testing_reference}
We first test the ability of CRG to model the reference galaxy with extreme color gradients from V- and I-band HST-like images of the galaxy. 
In particular, we investigate two potential limitations: 
\begin{enumerate}
    \item Impact of imperfect SEDs: For real images where the SEDs of the galaxies are unknown, we approximate the SEDs as linear. 
    We test the impact of these approximate SEDs by comparing the estimated CG bias when the CRG is modelled with linear SEDs and with true SEDs.
    \item Impact of pixel noise: The simulations analyzed in \secref{cg_para} were noise free. However, real galaxy images include pixel noise. We test the impact of this noise by comparing CG bias estimates from noisy and noise-free HST-like images.
\end{enumerate}

\begin{figure}[!ht]
\centering\includegraphics[width=0.75\linewidth]{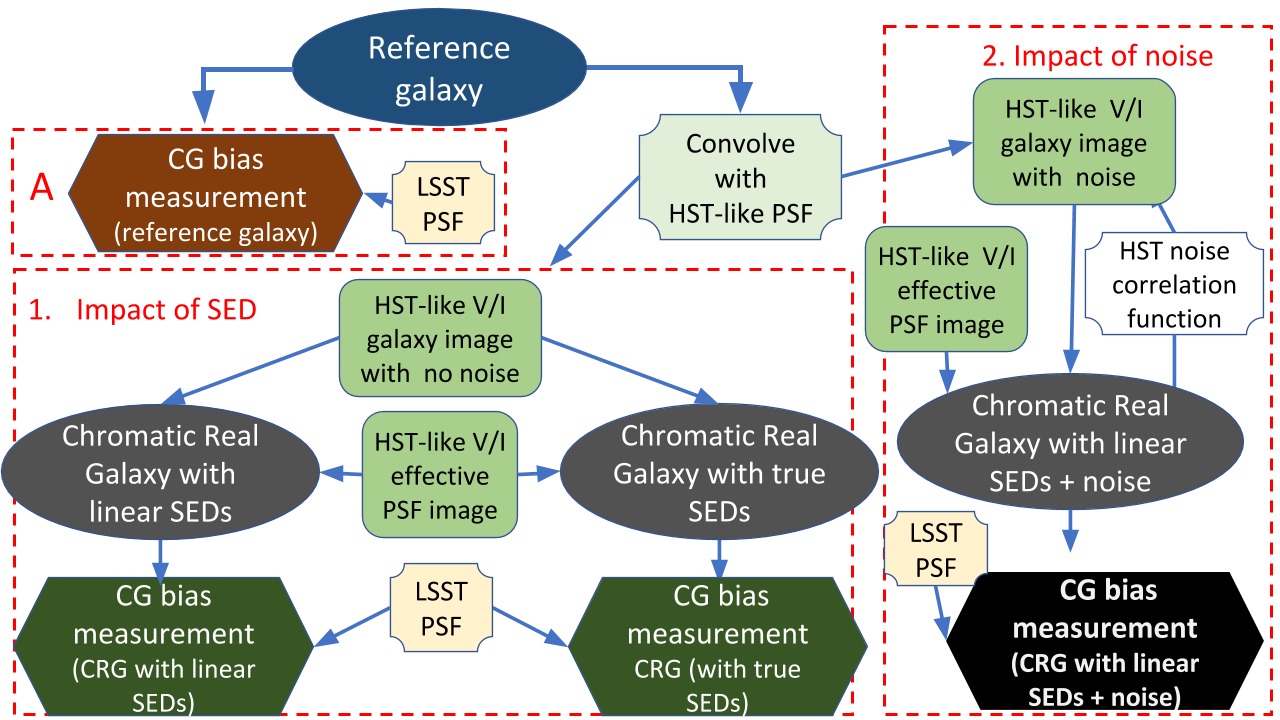}
\caption{Flow chart describing the method for testing the impact of (1) the linear approximation for the SED and (2) pixel noise, on the CRG model using HST-like images of the reference parametric galaxy with extreme color gradients. The CG bias estimated from 1 and 2 are then compared to the estimates from noise-free parametric simulations in A. See \appref{testing_reference} in text for details.}
\label{fig:ref_gal_flow}
\end{figure}

A flowchart describing the CRG test with the reference galaxy with extreme color gradients is shown in \figref{ref_gal_flow}. The red dashed box A shows the method used in \secref{cg_para} to estimate CG bias of the reference parametric galaxy when seen by LSST. 
Boxes 1 and 2 show the CRG tests to study the impact of imperfect SEDs and pixel noise, respectively. 

For Box 1, the HST-like images of the reference galaxy in V and I bands are input to CRG along with the effective HST PSF images. Two chromatic SBP of the galaxy are produced: one for CRG modelled with the true bulge and disk SEDs and the other with linear SEDs. Each of the modelled CRG profiles are then used to estimate CG bias similar to Box A.

We study the impact of noise in the input HST images (Box 2) by drawing the HST-like reference galaxy images with pixel noise in V and I bands. These noisy images are then input to CRG to model the chromatic SBP with linear SEDs. The chromatic SBPs are then used to estimate CG shear bias. Since the analysis would likely depend on the galaxy signal-to-noise ratio (SNR), we perform the study for varying noise levels for the input HST-like images.

\subsubsection{Parameters of HST-like images}
The bulge and disk SBPs have the same parameters as described in \secref{ref_param}. 
The HST and LSST parameters used in the simulations are summarized in \tabref{hst_lsst_parameters}. 
The HST-like images are drawn with a pixel scale of 0.03 arcsec and the LSST images with pixel scale 0.2 arcsec. The chromatic PSFs are assumed to have a Gaussian profile with a wavelength-dependent size as described in \eqnref{psf_sig}. The chromatic HST PSF is assumed to be diffraction-limited with wavelength-dependent size scaling exponent $\alpha=+1.0$. 
The PSF size at a reference wavelength of $\lambda^o=806$\,nm is set equal to $\sigma^o=0.071\rm \,arcsec$, which is the mean value from fits to real I-band HST PSF images.
The HST noise is modelled using the same noise correlation function described in \secref{aeg_charac}.

\begin {table*}[th]
\caption {HST and LSST parameters.} 
\label{tab:hst_lsst_parameters} 
\begin{center}
\begin{tabular}{ccccccc}
\hline
\hline
Survey  & Observing bands & $\alpha$ & PSF $\sigma^o$ & PSF $\lambda^o$ & Pixel size & Collecting area\\
 & &  & (arcsec) &  (nm) & (arcsec) & (m$^2$) \\
\hline
\hline
HST  & V/I   &$+1.0$ & 0.071 & 806 & 0.03 & 4.44 \\
LSST & $r/i$ &$-0.2$ & 0.297 & 550 &  0.2  & 32.4 \\
\hline
\end{tabular}
\end{center}
\textbf{Note.} The PSF is modelled as a circular Gaussian with size $\sigma^o$ at reference wavelength $\lambda^o$. 
The wavelength-dependent PSF size is determined by the scaling exponent $\alpha$ (see \eqnref{psf_sig}).
\end{table*}
 
\subsubsection{Results of CRG tests with reference galaxy with extreme color gradients}
\begin{figure*}[!ht]
\centering\includegraphics[width=0.8\linewidth]{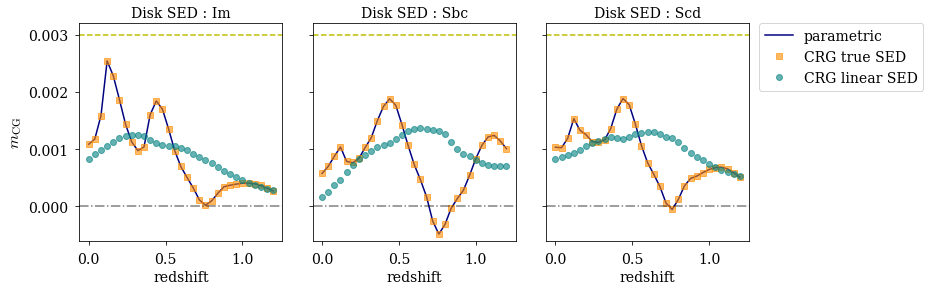}
\caption{Multiplicative shear bias due to CG, $m_{\rm CG}$, as a function of redshift for a reference galaxy with CWW-E bulge spectrum and three different disk SEDs (from left to right: CWW-Im, CWW-Sbc, CWW-Scd), for three types of noise-free simulations: parametric (solid blue line), CRG with the true SED (orange squares), and CRG with linear SEDs (teal dots). The yellow dashed line shows the LSST DESC requirement of 0.003 on the total systematic uncertainty in the redshift-dependent shear calibration.}
\label{fig:ref_gal_CG}
\end{figure*}

In \figref{ref_gal_CG} we show the estimated multiplicative shear bias due to CG, $m_{\rm CG}$, as a function of redshift for the reference galaxy with the three different disk SEDs (CWW-Im, CWW-Sbc, CWW-Scd) and a common CWW-E bulge SED.
The estimated bias $m_{\rm CG}$ from parametric simulations is shown as the solid blue line, identical to \figref{ref_gal_diff_sed}. 
We show the bias values measured by modeling the reference galaxy using {\tt ChromaticRealGalaxy} with the true SED (orange squares) and with a linear SED (teal dots). 
As expected, the biases estimated for the parametric simulation are in excellent agreement with the CRG-modelled SBP with true SEDs. 
However, the predicted biases for CRG with linear SEDs differ from these values, revealing the limitations in using zero and first-degree polynomials to reproduce the nonlinear features in the bulge and disk SEDs. 
In particular, CRG with a linear SED tends to smooth the variation in bias with redshift.
The difference between the redshift-averaged CG bias estimated with the linear SED and with the true SED is $\mathcal{O}(10^{-4}$), and the estimated biases satisfy the LSST DESC requirement of 0.003 on the total systematic uncertainty in the redshift-dependent shear calibration (yellow dashed lines) for both the linear and true SED models, for the three disk SEDs.
However, the total variation in CG bias over the redshift range $[0,1.2]$ is a factor of 2.6, 1.9 and 2.5 less when calculated with the linear SED rather than the true SED, for the CWW-Im, CWW-Sbc and CWW-Scd disk SEDs, respectively.

To assess the impact of noise we simulate HST-like images for the reference galaxy SBP with E-type bulge and Im-type disk SED, at a redshift of 0.3, but with different noise levels. 
As illustrated in Box 3 of \figref{ref_gal_flow}, we add HST-like noise to the simulated galaxy images before they are modelled with CRG. Since our noise-free analysis already showed that the predicted biases for CRG with true SEDs are identical to those for the parametric simulations, we perform the noise study for CRG with linear SEDs only. Since our ultimate aim is to model AEGIS galaxies with CRG, we add noise levels similar to those in the AEGIS images. 

\begin{figure}[!htpb]
\centering\includegraphics[width=.7\linewidth]{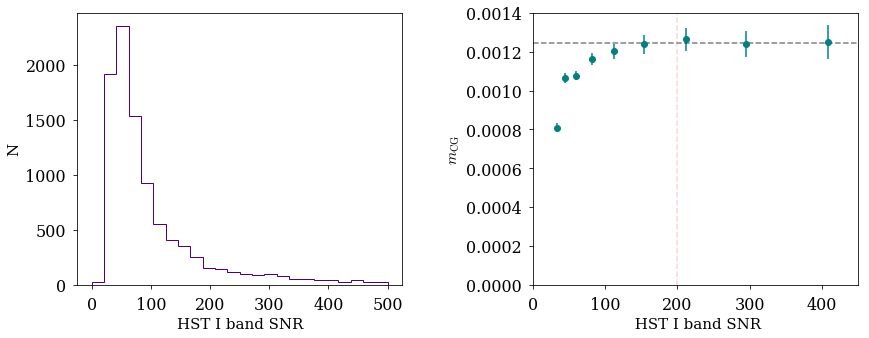}
\caption{Left: Distribution of I-band SNR of 10,000 randomly selected galaxies from the AEGIS catalog. Right: CG shear bias as a function of I-band SNR of the simulated HST-like images of the reference galaxy with extreme color gradients at a redshift of 0.3, modelled with CRG. The points correspond to the mean $m_{\rm CG}$ in each SNR bin and error bars show the statistical uncertainty on the mean. The black dashed line corresponds to the value of $m_{\rm CG}$ computed from the noise-free image of the galaxy modelled by CRG with linear SEDs ($m_{\rm CG} = 1.24 \times 10^{-3}$).}
\label{fig:ref_gal_cg_snr}
\end{figure}

The noisy reference galaxy is simulated 10,000 times, each with a different correlated noise realization to obtain an SNR level drawn randomly from the AEGIS catalog.
The sampled I-band SNR values are shown in the left panel of \figref{ref_gal_cg_snr}. 
The CG bias is then estimated using the CRG model of each of the noisy images.

The measured biases are binned by SNR in the right panel of \figref{ref_gal_cg_snr}. 
The solid dots denote the mean binned values of the measured CG bias and the error bars correspond to the statistical uncertainty on the means. The estimated CG bias shows a dependence on SNR even though all points correspond to the same galaxy SBP and thus have the same intrinsic color gradient.
The dashed line at $+0.00124$ corresponds to the estimated CG bias for the reference galaxy at a redshift of 0.3 drawn with no noise and modelled by CRG with linear SEDs. 
At high SNR, the mean value of the $m_{\rm CG}$ estimate approaches this noise-free CG bias. However, at low SNR the estimate diverges from the noise-free CG bias indicating that our methodology to isolate CG bias is not robust against noise bias. 
Hence we are limited to estimating CG bias for real galaxies with HST I-band SNR greater than 200. 

\subsection{Testing CRG with \textsc{CatSim} galaxies}
\label{app:crg_catsim}
We extend the CRG tests in the presence of noise to an ensemble of galaxies by repeating our analysis in Box 2 of \figref{ref_gal_flow} to galaxies in the \textsc{CatSim} catalog with a range of bulge and disk parameters. Since we concluded above that our methodology is applicable to only high SNR AEGIS galaxies, we compute the SNR of the HST-like galaxy as well as the SNR at 10-year LSST depth. 

\begin{figure}[!ht]
\centering\includegraphics[width=0.75\linewidth]{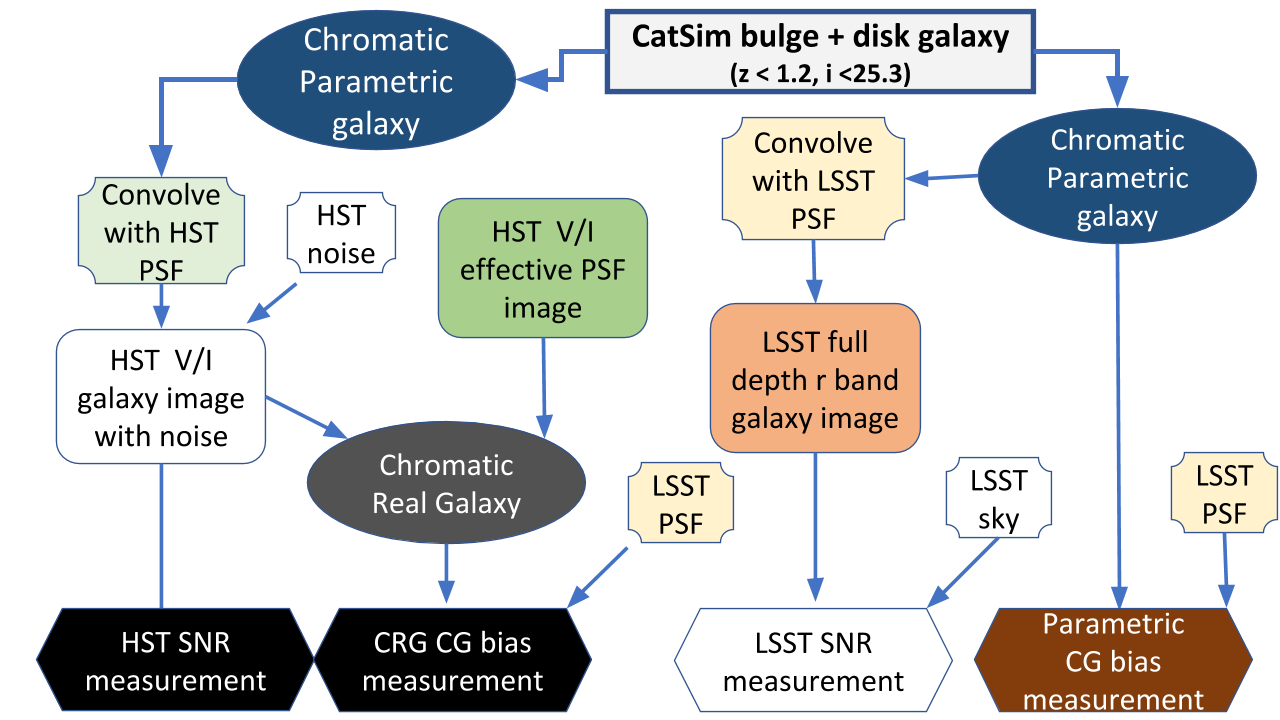}
\caption{Flow chart describing the method used to test CRG by measuring CG shear bias with galaxies parameterized in \textsc{CatSim}. See \appref{crg_catsim} in text for details.}
\label{fig:cat_flow}
\end{figure}

A flowchart describing the procedure is shown in \figref{cat_flow}. The left side of the flow chart illustrates the method used in \secref{cg_catsim} to compute CG bias for the parametric \textsc{CatSim} galaxies as measured by LSST and does not involve modelling with CRG. We also draw the galaxies in the $r$ band with sky noise corresponding to the 10-year LSST depth and compute their SNRs. 

The right side of the flow chart (in green) illustrates how we test CRG with HST-like \textsc{CatSim} galaxy images as input. The V- and I-band galaxy images are 
input to CRG, along with the HST effective PSF image and the correlation function for pixel noise, to model a chromatic SBP. CG bias is then estimated for these galaxies when observed in LSST $r$ and $i$ bands. For each noisy HST-like \textsc{CatSim} galaxy image, the V- and I-band SNRs are also computed. 

\subsubsection{Estimating SNR of \textsc{CatSim} galaxies}
\begin{figure}[!ht]
\centering\includegraphics[width=0.9\linewidth]{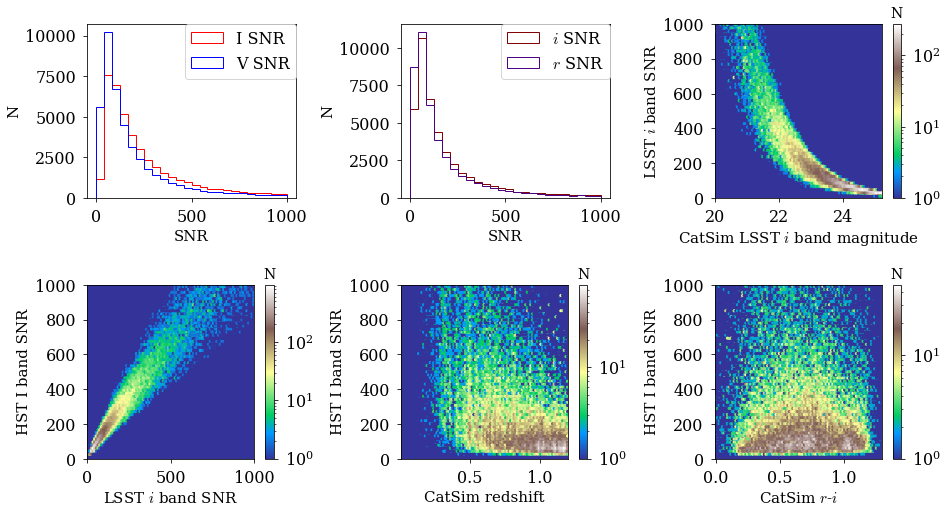}
\caption{HST and LSST SNR distributions for \textsc{CatSim} galaxies.
Histograms of SNR for HST V and I bands (top left) and LSST $r$ and $i$ bands (top center).
Calculated LSST $i$-band SNR versus $i$-band magnitude (top right), and HST I-band SNR versus LSST $i$-band SNR (bottom left). 
HST I-band SNR versus redshift (bottom center) and LSST $r-i$ color (bottom right).}
\label{fig:cat_gal_snr}
\end{figure}

The results of the SNR calculations for the galaxies are shown in \figref{cat_gal_snr}. The top panels show the computed SNR for the \textsc{CatSim} galaxies when seen by HST V and I bands (left) and LSST $r$ and $i$ bands (center). 
53\% of the galaxies have I-band SNR less than 200.
We plot the measured LSST $i$-band SNR against the true $i$-band \textsc{CatSim} catalog magnitude in the top right panel; the lower magnitude (i.e., brighter) galaxies have higher SNR as expected. 
In the bottom left plot, we compare HST I-band SNR and LSST $i$-band SNR. 
We do not expect the SNR in the two bands to be the same since the bandpasses, exposure times, collecting areas, and noise levels are different (see \tabref{hst_lsst_parameters}). 
However, since we have added a constant noise to all the HST and LSST simulations individually, we do expect the two SNRs to be linearly related. We observe that LSST $i$-band SNR is $\approx 0.7 \times$ HST I-band SNR. 
A selection criterion of I-band SNR $>$ 200 for the AEGIS galaxies effectively eliminates all galaxies with LSST $i$-band SNR less than $\approx$ 80.
 
In the final two panels, we show the correlation between I-band SNR and redshift (center) and $r-i$ color (right). The magnitudes, redshifts and colors are from the \textsc{CatSim} catalog and are noise free. 
High redshift galaxies have, on average, smaller SNRs, which is expected since the higher redshift galaxies tend to have lower observed flux. As a consequence, our CG bias estimates from AEGIS galaxies are not as reliable for higher redshift galaxies.
There appears to be a weak correlation between SNR and color. 
 
\subsubsection{CG shear bias estimates from noisy \textsc{CatSim} galaxies with CRG}
\begin{figure}[!ht]
\centering\includegraphics[width=.6\linewidth]{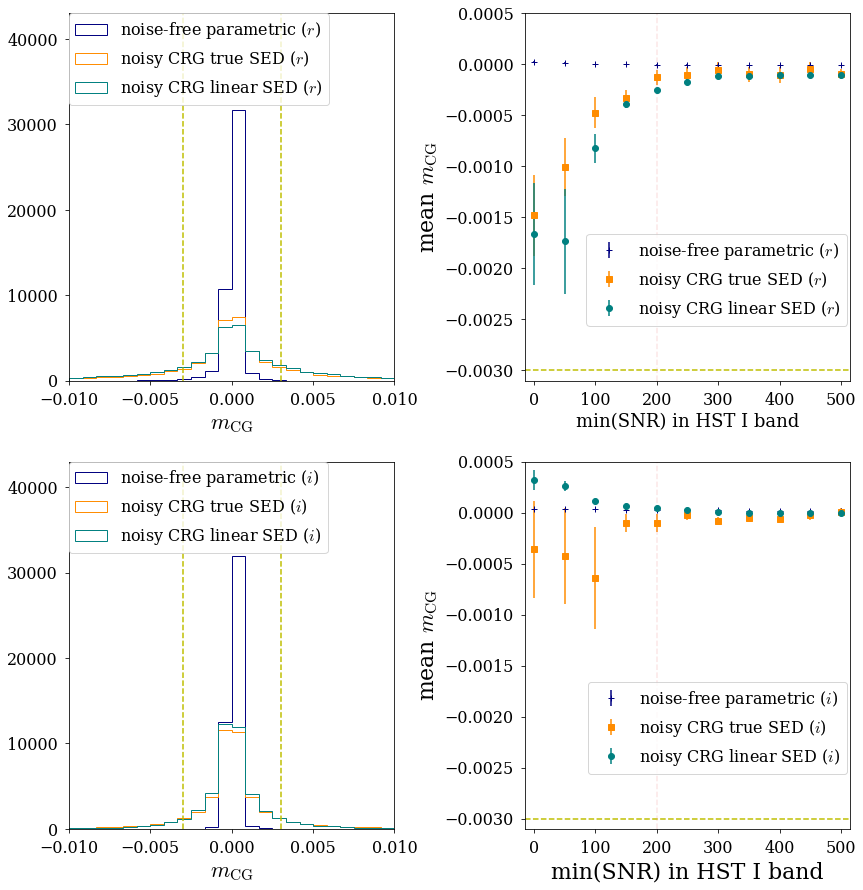}
\caption{Left panels: Distribution of $m_{\rm CG}$ in the $r$ band (top) and $i$ band (bottom) for \textsc{CatSim} galaxies. Right panels: Mean $m_{\rm CG}$ in the $r$ band (top) and $i$ band (bottom) for galaxies with I-band SNR $>$ min(SNR). Results from parametric simulations are shown for true SEDs in blue, for CRG with linear SEDs in teal, and CRG with true SED input in orange. The yellow dashed lines show the LSST DESC requirement of 0.003 on the total systematic uncertainty in the redshift-dependent shear calibration.}
\label{fig:cat_mCG_i_r}
\end{figure}

\figref{cat_mCG_i_r} plots the results from the CG bias estimates for all the noisy HST-like \textsc{CatSim} galaxies modelled by CRG and observed in LSST $r$ band (top) and $i$ band (bottom). 
For reference we also show the estimates from noise-free parametric simulations (blue). 
The panels on the left show histograms of the estimated CG bias from noisy HST-like \textsc{CatSim} images modelled by CRG with true SEDs (orange) and linear SEDs (teal). While the noise-free parametric galaxies exhibit small biases, the estimates from noisy CRG-modelled images show large tails due to contributions from pixel noise. This is more evident in the right panels where we plot the mean CG bias estimates for galaxies with SNR higher than min(SNR) in the HST I band. Noise impacts the CG bias estimates for both linear and true SED models at low SNR. With no SNR cut on the HST-like images, the mean CG bias estimates from CRG galaxies modelled with linear SEDs in $r$ band is $m_{\rm CG} = (-1.7 \pm 0.5)\times 10^{-3}$ while the median of the distribution is $-4.1 \times 10^{-6}$ with a median absolute deviation (MAD) of $3.3 \times 10^{-3}$. The mean and median bias estimates in the $i$ band are $3.2 \times 10^{-4}$ and $4.1 \times 10^{-6}$, respectively.

The pixel noise causes the bias estimated from HST-like galaxies modelled by CRG with true SEDs to be more negative in both bands, while the impact of imperfect SEDs causes the bias to be more negative in the $r$ band and positive in the $i$ band. 
The impact of noise bias on the CG bias estimates is less pronounced in the $i$ band. 
This is because the galaxy images have higher SNR in $i$ band than $r$ band as shown in \figref{cat_gal_snr}. As more low-SNR galaxies are excluded, the bias measured with the CRG method approaches the value determined with the parametric analysis.
For galaxies with HST I-band SNR $>$ 200, the mean color gradient bias is $m_{\rm CG} = (-2.5 \pm 0.3) \times{10^{-4}}$ in the $r$ band and $m_{\rm CG} = (0.47 \pm 0.10) \times{10^{-4}}$ in the $i$ band.
The LSST requirement on redshift dependence of the shear bias from all systematic effects is shown as the dashed yellow line. 

\begin{figure*}[!ht]
\centering\includegraphics[width=.8\linewidth]{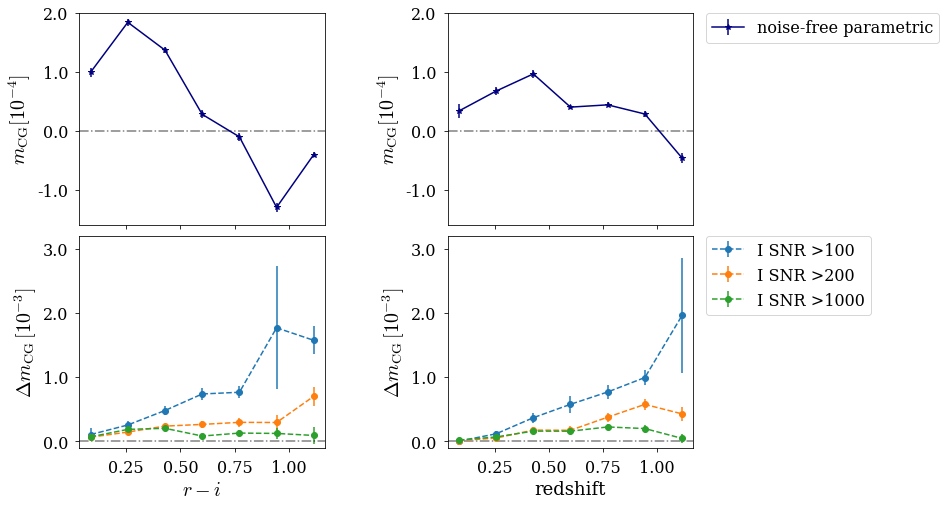}
\caption{Multiplicative shear bias from CG, $m_{\rm CG}$, as a function of $r-i$ color (left) and redshift (right) for \textsc{CatSim} galaxies.
Top: $m_{\rm CG}$ estimates from noise-free parametric simulations.
Bottom: Difference in mean $m_{\rm CG}$ estimated from noise-free parametric simulations and HST-like galaxies modelled with CRG ($\Delta m_{\rm CG}$) for three different HST I-band SNR cutoffs. The error bars correspond to the statistical uncertainties on the binned means.}
\label{fig:cat_mCG_bin}
\end{figure*}

Since the ability of {\tt ChromaticRealGalaxy} to model chromatic features and accurately estimate the magnitude of CG bias depends on the intrinsic properties of the galaxies, we investigate the dependence of the estimated CG bias on the galaxy color and redshift. In \figref{cat_mCG_bin}, we show the dependence of estimated CG bias on $r-i$ color (left) and redshift (right). 
The top plots correspond to noise-free parametric simulations of \textsc{CatSim} galaxies. 
The bottom plots illustrate the impact of applying three different minimum SNR cutoffs. 
The dots denote the difference in mean $m_{\rm CG}$ for noise-free simulations (top panel) and mean $m_{\rm CG}$ for HST-like galaxies modelled with CRG ($\Delta m_{\rm CG}$), with HST I-band SNR $>$100, 200 and 1000.
The difference in the mean bias decreases with increasing SNR cutoffs.
Selection effects due to the SNR cutoffs are found to bias the binned means of CG bias estimates for noise-free parametric simulations by $\mathcal{O}(10^{-4})$. 
Therefore, our method of using {\tt ChromaticRealGalaxy} to model galaxies observed by LSST from noisy HST images is able to reproduce CG bias with an accuracy of $\mathcal{O}(10^{-4})$ for galaxies with HST I-band SNR$>$200.

\end{appendix}
\end{document}